\documentstyle[amssymb,12pt]{amsart}
\newcommand{\N}{{\Bbb N}}
\newcommand{\Z}{{\Bbb Z}}
\newcommand{\C}{{\Bbb C}}

\newcommand{\Hp}{\mbox{${\operatorname H}_{+}$}}
\newcommand{\Lg}{L{\g}^{\wedge}}
\newcommand{\Vh}{V^{\wedge}}
\newcommand{\Vhn}{V^{\wedge[n]}}
\newcommand{\Vn}{V^{[n]}}

\newcommand{\dt}{{\operatorname d}t}

\newcommand{\zon}{z}

\newcommand{\gz}{\Lie(z_1,\dots,z_n,\tau,\lambda)}
\newcommand{\Ref}[1]{{\rom{(}\ref{#1}\rom{)}}}
\newcommand{\bean}{\begin{eqnarray}}
\newcommand{\eean}{\end{eqnarray}}
\newcommand{\be}{\begin{displaymath}}
\newcommand{\ee}{\end{displaymath}}
\newcommand{\bea}{\begin{eqnarray*}}
\newcommand{\eea}{\end{eqnarray*}}

\newcommand{\dee}{{\operatorname{d}}}
\newcommand{\ddt}{\frac d{dt}}
\newcommand{\co}{C^{[n]}}
\newcommand{\Lie}{\cal L}
\newcommand{\Gr}{\operatorname{Gr}}
\newcommand{\g}{{{\frak g}\,}}
\newcommand{\No}{m}
\newcommand{\bor}{{{\frak b}}}

\newcommand{\h}{{{\frak h\,}}}
\newcommand{\rank}{\operatorname{rank}}
\newcommand{\ad}{\operatorname{ad}}
\newcommand{\Id}{\operatorname{Id}}
\newcommand{\Ad}{{\operatorname{Ad}}}
\newcommand{\half}{\frac12}
\newcommand{\hor}{{\operatorname{hor}}}
\newcommand{\SL}{\operatorname{SL}}
\newcommand{\vs}{\vspace{.5\baselineskip}}
\newcommand{\Eta}{\operatorname{H}}
\newcommand{\reg}{{ss}}
\newenvironment{proof}{\noindent{\em Proof\/}:}{$\;\Box$}
\newenvironment{definition}{\par\vspace{.5\baselineskip}
\noindent{\em Definition\/}:}{\par\vspace{.5\baselineskip}}
\newtheorem%
{thm}{Theorem}[section]
\newtheorem%
{proposition}[thm]{Proposition}
\newtheorem%
{lemma}[thm]{Lemma}
\newtheorem%
{corollary}[thm]{Corollary}
\newcommand{\parag}[1]{\subsection{{#1}}}
\title[Conformal blocks on elliptic curves]
{Conformal blocks on elliptic curves and
the Knizhnik--Zamolodchikov--Bernard
equations}
\author{Giovanni Felder}
\thanks{The first author was supported
in part by NSF grant DMS-9400841}
\address{G. F.: Department of Mathematics\\
University of North Carolina at Chapel Hill\\
Chapel Hill, NC 27599-3250\\
USA}
\author{Christian Wieczerkowski}
\address{C. W.: Institut f\"ur theoretische Physik I\\
Universit\"at M\"unster\\ Wilhelm-Klemm-Str.\ 9\\
D-48149 M\"unster\\
Germany}
\begin{document}
\maketitle
\begin{abstract}
We give an explicit description of the vector bundle
of WZW conformal blocks on elliptic curves with marked points
as subbundle of a vector bundle of Weyl group invariant
vector valued theta functions on a
Cartan subalgebra. We give a partly conjectural characterization
of this subbundle in terms of certain vanishing conditions
on affine hyperplanes. In some cases, explicit calculation
are possible and confirm the conjecture.
The Friedan--Shenker flat connection
is calculated, and it is shown that horizontal sections
are solutions of Bernard's generalization of the Knizhnik--%
Zamolodchikov equation.
\end{abstract}

\section{Introduction}
\noindent The aim of this work is to give a description
of conformal blocks of the Wess--Zumino--Witten model
on genus one curves as explicit as on the Riemann sphere.

Let us recall the well-known situation on the sphere. One fixes a
simple finite dimensional complex Lie algebra $\g$, with invariant
bilinear form $(\ ,\ )$
normalized so that the longest roots have length squared
2, and a positive integer $k$ called level.  One then considers the
corresponding affine Kac--Moody Lie algebra, the one dimensional
central extension of the loop algebra $\g\otimes\C((t))$ associated
to the 2-cocycle $c(X\otimes f,Y\otimes g)=(X,Y)\operatorname{res}dfg$.  Its
irreducible highest weight integrable representations of level (=
value of central generator) $k$ are in one to one correspondence with
a certain finite set $I_k$ of finite dimensional irreducible
representations of $\g$. These representations extend, by the Sugawara
construction, to representations of the affine algebra to which an
element $L_{-1}$ is adjoined, such that
 $[L_{-1},X\otimes f]=-X\otimes\frac d{dt}f$. Then to
each $n$-tuple of distinct points $z_1$, \dots, $z_n$ on the complex
plane, and of representations $V_1$, \dots, $V_n$ in $I_k$ one
associates the space of conformal blocks $E(z_1,\dots,z_n)$. It is the
space of linear forms on the tensor product $\otimes_1^n
V_i^\wedge$ of the corresponding level $k$ representations of the
affine algebra, which are annihilated by the Lie algebra
$\Lie(z_1,\dots,z_n)$ of $\g$-valued meromorphic functions with poles
in $\{z_1,\dots,z_n\}$ and regular at infinity.  The latter algebra
acts on $\otimes V_i^\wedge$ by viewing $\Lie(z_1,\dots,z_n)$ as a Lie
subalgebra of the direct sum of $n$ copies of the loop algebra via
Laurent expansion at the poles.  The central extension does not cause
problems as the corresponding cocycle vanishes on $\Lie(z_1,
\dots,z_n)$ in virtue of the residue theorem.

It turns out that the spaces $E(z_1,\dots,z_n)$ are
finite dimensional and are the
fibers of a holomorphic vector bundle over the configuration
space $\C^n-$ diagonals, carrying the flat connection
$d-\sum_idz_iL_{-1}^{(i)}$ ($L_{-1}^{(i)}$
 acts on the right of a linear form)
given in terms of the Sugawara construction.
We use the notation $X^{(i)}=\cdots\otimes\Id
\otimes X\otimes\Id\cdots$ to denote the action of
 on the $i$th factor of a tensor product.

This part of the construction generalizes to surfaces of arbitrary
genus (see \cite{TUY}). What is new is that one has to also specify local
coordinates around the points $z_i$ to give a meaning to the Laurent
expansion, and that the connection is in general only projectively
flat (i.e., the curvature is a multiple of the identity).

To give a more explicit description of the vector bundle of
conformal blocks on the sphere, and in particular to compute
the holonomy of the connection, one observes that the
map $E(z_1,\dots,z_n)\to (\otimes_iV_i)^*$ given by
restriction to $V_i\subset V_i^\wedge$ is injective. Thus
we can view $E$ as a subbundle of a trivial vector bundle
of finite rank. This subbundle can be described by an explicit
algebraic condition \cite{FeScVa}. After this identification
the connection can be given in explicit terms and the equation
for horizontal sections reduces to the famous Knizhnik--Zamolodchikov
equations
\be
(k+h^\vee)\partial_{z_i}\omega(z_1,\dots,z_n)
=\sum_{j:j\neq i}\sum_a
\frac{T_a^{(i)}T_a^{(j)}}{z_i-z_j}\omega(z_1,\dots,z_n).
\ee
In this equation, $h^\vee$ is the dual Coxeter number of $\g$ and
$T_a$, $a=1,\dots, \operatorname{dim}(\g)$
 is any orthonormal basis of $\g$. We view here the
dual spaces $V_i^*$ as contragradient representations.

Let us now consider the situation on genus one curves,
which we view as $\C/\Z+\tau\Z$ for $\tau$ in the upper
half plane. Let us denote by $E(z_1,\dots,z_n,\tau)$ the
space of conformal blocks. Again, by \cite{TUY}, this is
the finite dimensional fiber of a holomorphic vector
bundle with flat connection on the elliptic configuration
space $\co$ of $n+1$-tuples $(z_1,\dots,z_n,\tau)$ with
Im($\tau)>0$ and $z_i\neq z_j$ mod $\Z+\tau\Z$ if $i\neq j$.

The trouble is that the restriction to $\otimes_i V_i$ is
no longer injective, the reason being that there are no
meromorphic functions on elliptic curves with one simple
pole only. The way out is the following construction which
brings the moduli space of flat $G$-bundles
into the game. Consider the  Lie algebras
$\Lie(z_1,\dots,z_n,\tau,\lambda)$, parametrized by $\lambda$
in a Cartan subalgebra $\h$, of $\Z$-periodic meromorphic functions
$X:\C\to\g$ with poles at $z_1$, \dots, $z_n$ modulo
$\Z+\tau \Z$, such that $X(t+\tau)=\exp(2\pi i\ad \lambda)X(t)$.

These algebras act on $\otimes_iV_i^\wedge$ and we can define
a space of (twisted) conformal blocks $E_\h(z,\tau,\lambda)$ as
space of invariant linear forms (see \ref{tcb}). The original space of
conformal blocks is recovered by setting $\lambda=0$.

It turns out that $E_\h(z,\tau,\lambda)$ is again the
fiber over $(z,\tau,\lambda)$ of a holomorphic vector
bundle $E_\h$ over $\co\times\h$ with flat connection,
whose restriction to $\co\times\{0\}$ is $E$. Thus
we can by parallel transport in the direction of $\h$ identify
the space of sections $E(U)$ of $E$
over an open set $U\subset \co$
with the space of sections of $E_\h$ which are horizontal in
the  direction of $\h$:
\be
E(U)\simeq E_\h(U\times\h)^{\hor}.
\ee
The point is now that the restriction map
\be
E_\h(U\times\h)^\hor\to (\otimes_iV_i)^*\otimes\cal O(U\times\h),
\ee
to $\Vn$
is injective (Proposition \ref{iota}). Composing these two maps we may view
the vector bundle of conformal blocks as a subbundle
of an explicitly given vector bundle on $\co$ of finite rank.
 Indeed
we show (Theorems \ref{main}, \ref{main1})
 that the image is contained in the space of
functions on $U\times\h$ which have definite transformation
properties (of theta function type)
under translations of $\lambda$ by $Q^\vee+\tau Q^\vee$
where $Q^\vee$ denotes the coroot lattice. Moreover
the theta functions in the image are invariant under a natural
action of the Weyl group, and obey a certain vanishing
condition as the argument approaches affine root hyperplanes.
We conjecture that these conditions characterize
completely the image. This conjecture is confirmed
in some cases, including
a special case which arises \cite{EtKi} in the
theory of quantum integrable many body problems (see
\ref{kzbe}): we describe explicitly the space of conformal
blocks in the case of $sl_N$, $n=1$, where the
representation is any symmetric power of the defining
$N$-dimensional representation.

The characterization of conformal blocks in terms
of invariant theta functions obeying vanishing conditions
was first given (in the $sl_2$ case) by Falceto and Gaw\c edzki
\cite{FaGa}, who define conformal blocks as Chern--Simons
states in geometric quantization.

After the identification of conformal blocks as subbundle
of the ``invariant theta function'' bundle, we describe the
connection in explicit terms (Theorem \ref{KZB}),
and get a generalization
of the Knizhnik--Zamolodchikov equations. These equations
were essentially written by Bernard \cite{Be1}, \cite{Be2}
in a in a slightly different context, and were recently
reconsidered from a more geometrical point of view in
\cite{FaGa}. They have the form (see Sect.\ \ref{kZb})
\begin{eqnarray*}
\kappa\partial_{z_j}\tilde\omega&=&
-\sum_\nu h_\nu^{(j)}\partial_{\lambda_\nu}\tilde\omega
+\sum_{l:l\neq j}
\Omega^{(j,l)}(z_j-z_l,\tau,\lambda)\tilde\omega,\\
4\pi i\kappa\partial_\tau\tilde\omega&=&
\sum_\nu\partial_{\lambda_\nu}^2\tilde\omega
+\sum_{j,l}
\Eta^{(j,l)}(z_j-z_l,\tau,\lambda)\tilde\omega,
\end{eqnarray*}
for some tensors $\Omega$, $\Eta\in\g\otimes\g$, given in terms
of Jacobi theta functions. Here $\tilde\omega$ is related
to $\omega$ by multiplication by the Weyl--Kac denominator.
Thus, the right way to look at these equations is to view
$u$ as a section of a subbundle of the vector bundle over the
elliptic configuration
space of $n+1$ tuples $(z_1,\dots,z_n,\tau)$, whose fiber
is a finite dimensional space of invariant theta functions.

In this paper we do not discuss an alternative approach to conformal
blocks on elliptic curves, which is in terms of traces of products of
vertex operators. Bernard \cite{Be1} showed that such traces obey his
differential equations. Using this formulation, integral representation
of solutions were given in the $sl_2$ case in \cite{BeFe}. To complete
the picture, one should show that those solutions are indeed theta
functions with vanishing condition.

Let us also point out the recent paper \cite{EtFrKi} that shows that the
same space of invariant theta functions with vanishing condition can be
identified with a space of equivariant functions on the corresponding
Kac--Moody group.

Some of the results presented here were announced in
\cite{FeWi}.

\medskip
\noindent{\sc Acknowledgments.}
We wish to thank K. Gaw\c e\-dzki, S. Kumar and A. Varchenko
for useful comments and
discussions. We thank H. Wenzl for teaching us how
to compute Verlinde dimensions.
 Parts of this work
were done while the first author was visiting
the Isaac Newton Institute of Mathematical Sciences
(September 1992), the Institut des Hautes Etudes
Scientifiques (Fall 1993), and the Forschungsinstitut
f\"ur Mathematik, ETH (Summer 1994). G. F. wishes to
thank these institutions for hospitality.

\section{Conformal blocks on elliptic curves}
\parag{Elliptic configuration spaces}
Let $\Hp=\{\tau\in\C\,|\,{\operatorname{Im}}\,\tau>0\}$ be the upper half
plane and for $\tau\in \Hp$ denote by $L(\tau)$ the
lattice $\Z+\tau\Z\subset\C$. Let $n$ be
a positive integer.
 We define the elliptic configuration space to be
the subset of $\C^n\times\Hp$ consisting
of points $(z_1,\dots,z_n,\tau)$ so that $z_i\neq z_j$
mod $L(\tau)$ if $i\neq j$.

The space of points $(z,\tau)\in \co$ with fixed
$\tau$ is a covering of the configuration space of
$n$ ordered points on the elliptic curve $\C/L(\tau)$.

\parag{Meromorphic Lie algebras}\label{Cas}
 Let  $\g$ be a complex simple Lie algebra
with dual Coxeter number $h^\vee$
and $k$ be a positive integer. Fix a Cartan subalgebra $\h$
of $\g$ and let $\g=\h\oplus(\oplus_{\alpha\in\Delta}\g_\alpha)$
be the corresponding root space decomposition. The invariant
bilinear form is normalized in such a way that $(\alpha^\vee,
\alpha^\vee)=2$ for long roots $\alpha$ (see \cite{Hu}).
We choose an orthonormal basis $(h_\nu)$ of $\h$. The
symmetric invariant tensor $C\in\g\otimes\g$ dual to $(\ ,\ )$
admits then a decomposition
$C=\sum_{\alpha\in\Delta\cup\{0\}}C_\alpha$,
with $C_0=\Sigma \,h_\nu\otimes h_\nu\in\h\otimes\h$ and
$C_\alpha\in\g_\alpha\otimes\g_{-\alpha}$, if $\alpha\in\Delta$.

We define a family of Lie algebras of meromorphic
functions with values in $\g$ parametrized by $\co\times\h$

\begin{definition}
For $(z,\tau)=(z_1,\dots,z_n,\tau)\in\co$ and $\lambda\in \h$, let
$\Lie(z,\tau,\lambda)$
be the Lie algebra of meromorphic functions $t\mapsto X(t)$
on the complex
plane with values in $\g$ such that
\be
X(t+1)=X(t),\qquad
X(t+\tau)=\exp(2\pi i \,\operatorname{ad} \lambda)X(t),
\ee
and whose poles belong to $\cup_{i=1}^nz_i+L(\tau)$.
More generally, for any open set $U\subset \co\times\h$
 let $\Lie_\h(U)$ be the Lie algebra
of meromorphic functions $(t,z,\tau,\lambda)\mapsto
X(t,z,\tau,\lambda)$ on $\C\times U$  with
values in $\g$,
whose poles are on the hyperplanes $t=z_i+r+s\tau$,
$1\leq i\leq n$, $r,s\in\Z$, and such that for all
$(z,\tau,\lambda)\in U$, the function
$t\to X(t,z,\tau,\lambda)$ belongs to $\Lie(z,\tau,\lambda)$.
Similarly, define $\Lie(U)$ for an open subset $U$ of
$\co$ to be the Lie algebra of
meromorphic functions $(t,z,\tau)\mapsto
X(t,z,\tau)$ on $\C\times U$  with
values in $\g$,
whose poles are on the same hyperplanes,
 and such that for all
$(z,\tau)\in U$, the function
$t\to X(t,z,\tau)$ belongs to $\Lie(z,\tau,0)$.
\end{definition}

\noindent An explicit description of these Lie algebras is given in
Appendix \ref{A}. An important property is that they have
a filtration by locally free finitely generated  sheaves:
Let $\cal O(U)$ be the algebra of holomorphic functions
on an open set $U\subset \co\times \h$, and for any non-negative
integer $j$ let $\Lie_\h^{\leq j}(U)$ be the $\cal O(U)$-submodule
of $\Lie_\h(U)$
consisting of functions whose poles have order at most $j$.
Similarly we define $\Lie^{\leq j}(U)$ for open sets $U\in\co$.
The assignments $U\to\Lie^{\leq j}(U)$,
 $U\to\Lie_\h^{\leq j}(U)$ are  sheaves of $\cal O$-%
modules.

\begin{proposition}\label{free} $\Lie_\h^{\leq j}$ is a locally free,
locally finitely generated sheaf of  $\cal O$-mod\-ules.
In other words, every point in $\co\times\h$ has a
neighborhood $U$ such that $\Lie_\h^{\leq j}(U)\simeq
\C^{n_j}\otimes\cal O(U)$ as an $\cal O(U)$-module,
for some $n_j$. Moreover for each $x\in \co\times\h$, every
$X\in\Lie(x)$ extends to a function in $\Lie_\h^{\leq j}(U)$
for some $j$ and $U\ni x$. The same results hold
for $\Lie^{\leq j}$.
\end{proposition}

\noindent
The proof is contained in Appendix \ref{A} (see Corollary \ref{freeA}).

\parag{Tensor product of affine Kac--Moody algebra modules}
\label{tcb}
Let $L\g=\g\otimes\C((t))$ be the loop algebra of $\g$.
Fix a positive integer $k\in\N$.
Let $\Lg=L\g\oplus\C K$ be the central extension of $L\g$
associated with the 2-cocycle
\be
c(X\otimes f,Y\otimes g)=
(X,Y)\,{\operatorname{res}}(f^\prime g \dt),
\ee
where the residue of a formal Laurent series is given by
${\operatorname{res}}(\sum_n a_n t^n\dt)=a_{-1}$. Thus the Lie bracket in
$\Lg$ has the form
\be
[X\otimes f\oplus\zeta K, Y\otimes g\oplus\xi K]=
[X,Y]\otimes fg \oplus c(X,Y)K.
\ee
With every irreducible highest weight $\g$-module $V$
is associated an irreducible highest weight $\Lg$-module $\Vh$ of
level $k$.
Its construction goes as follows.
The action of $\g$ is first extended to the Lie  subalgebra
$\bor_+=\g\otimes\C [[t]]\oplus\C K$ of $\Lg$, by letting
$\g\otimes t\C[[t]]$ act by zero and the central element
$K$ by $k$.
Then a generalized Verma module
$\tilde V=U(\Lg)\otimes_{U({\rm b}_+)}V$
is induced. It is freely generated by (any basis of) $V$
as a $\g\otimes t^{-1}\C[t^{-1}]$-module.
The polynomial subalgebra $L\g_P=\g\otimes\C[t,t^{-1}]\oplus \C$ of
$\Lg$ is $\Z$-graded with deg$(X\otimes t^j)=-j$.
Since $\tilde V\simeq U(\Lg_P)\otimes_{U(\bor_+\cap \Lg_P)}V$,
the generalized Verma module is naturally graded.
By definition
the irreducible module $\Vh$ is the quotient of the
generalized Verma module by its maximal proper
graded submodule.

We will consider integrable modules,  $\Vh$ of fixed
level $k=0,1,2,\dots$.  If we fix
a set of simple roots $\alpha_1,\dots,\alpha_l\in\Delta$,
and denote by $\theta$ the corresponding highest root,
$V^\wedge$ is integrable of level $k$
if the irreducible $\g$-module
$V$ has highest weight $\mu$ in the the subset
\begin{equation}
\label{ik}
I_k=\{\mu\in P|
(\mu,\alpha_i)\geq 0,\quad i=1,\dots,l,\quad
(\mu,\theta)\leq k\},
\end{equation}
of the weight lattice $P$.
Let $v$ be the highest weight vector of $V$ and
$e_\theta$ a generator of $\g_\theta$. Then the maximal proper submodule
is generated by $(e_\theta\otimes t^{-1})^{k-(\mu,\theta)+1}
v$.

The grading extends to $\Vhn$ by setting deg($v_1\otimes\cdots
\otimes v_n)=\Sigma\operatorname{deg}(v_i)$. With our convention
all homogeneous vectors have non-negative degree.

Fix $n$ highest weight $\g$-modules
$V_j$, $1\leq j\leq n$ such that the corresponding
$\Lg$-modules $V_j^\wedge$ are integrable of level $k$,
and let $\tau\in \Hp$ and
$z_1,\dots, z_n$ complex
numbers with $z_i\neq z_j$ mod $L(\tau)$ if $i\neq j$.
We think of $\Vh_j$ as an $\Lg$-module which is
attached to the point $z_j$.

In the following we will use the abbreviations
$V^{[n]}=V_1\otimes\cdots\otimes V_n$ and
$\Vhn=\Vh_1\otimes\cdots\otimes\Vh_n$.

We now construct an action of $\Lie(\zon,\tau,\lambda)$
on $\Vhn$.
For $X\in\Lie(\zon,\tau,\lambda)$ let
$\delta_j(X)=X(z_j+t)\in\g\otimes\C((t))$ be the Laurent expansion of $X$
at $z_j$ viewed as a formal Laurent series in $t$.
Then
\begin{equation}\label{PI}
\delta(X)=\delta_1(X)\oplus\cdots\oplus\delta_n(X),
\label{eq:pi(x)}
\end{equation}
defines a Lie algebra embedding of $\Lie(\zon,\tau,\lambda)$ into
$L\g\oplus\cdots\oplus L\g$.
As a vector space
$L\g\oplus\cdots\oplus L\g$ is embedded in
$\Lg\oplus\cdots\oplus\Lg$.
The embedding is of course not a Lie algebra homomorphism.
Since $\Lg\oplus\cdots\oplus\Lg$ acts on
$\Vhn$ we obtain a map from $\Lie(\zon,\tau,\lambda)$
to ${\operatorname{End}}_\C(\Vhn)$. This map will also be denoted by
$\delta$. Thanks to the residue theorem it turns out to be
a Lie algebra homomorphism.

\begin{proposition}
For $X$, $Y\in\Lie(\zon,\tau,\lambda)$,
\be
\delta([X,Y])=[\delta(X),\delta(Y)].
\ee
\end{proposition}
\begin{proof}
In ${\operatorname{End}}_{\C}(\Vhn)$ we have the equation
\be
\delta([X,Y])=
[\delta(X),\delta(Y)] +
k\sum_{j=1}^n {\operatorname{res}}_{t=z_j}
\left( (X^\prime(t),Y(t))\,dt \right).
\ee
But $(X^\prime(t),Y(t))$ is doubly periodic (by ad-invariance
of $(\ ,\ )$) so that
the sum of residues vanishes.
\end{proof}\par\medskip

\parag{Vector fields}
The Lie algebra $\operatorname{Vect}(S^1)=\C((t)){\dee\over\dee t}$ of formal
vector fields on the circle acts by derivations on
$L\g$. Let us denote this action
simply by $(\xi(t)\ddt,X(t))\mapsto\xi(t)\ddt X(t)$. It
extends to an action on $\Lg$ by letting vector fields
act trivially on the center.
The Sugawara construction
yields a projective representation of $\operatorname{Vect}(S^1)$ on
$V^\wedge$, for any finite dimensional $\g$-module $V$.
The Sugawara operators $L_n\in\operatorname{End}(V^\wedge)$
are defined by choosing any
basis $\{B_1,\dots,B_d\}$ of $\g$, with dual basis
$\{B^1,\dots,B^d\}$ of $\g$ so that $(B^a,B_b)=\delta_{ab}$,
and setting
\bea
L_n&=&{1\over 2(k+h^\vee)}
\sum_{m\in\Z}\sum_{a=1}^d(B^a\otimes t^{n-m})(B_a\otimes t^m),
\qquad n\neq 0
\\
L_0&=&\frac12[L_1,L_{-1}].
\eea
These operators are independent of the choice of
basis and obey the commutations relations
of the Virasoro algebra $[L_n,L_m]=(n-m)L_{n+m}+
\frac c{12}(n^3-n)$ with central charge
$c=k\operatorname{dim}(\g)/(k+h^\vee)$.
Then
\be
\sum_n\xi_nt^{n+1}
\ddt\mapsto -\sum_n\xi_nL_n\in\operatorname{End}(V^\wedge),
\ee
defines a projective representation of $\operatorname{Vect}(S^1)$
on $V^\wedge$, with the intertwining property
$[\xi(t)\ddt,X(t)]=\xi(t)\ddt X(t)$, for any $X(t)\in \Lg$.
Note that all infinite sums are actually
finite when acting on any vector in $V^\wedge$.

\parag{Conformal blocks}
For a Lie algebra module $V$ we denote
by $V^*$ the dual vector space with natural (right)
action of the Lie algebra. The notation $\langle\omega,v\rangle$
will be used to denote the evaluation of a linear form
$\omega$ on a vector $v$.

\begin{definition}
The space of {\em twisted conformal blocks}
associated to data $\g$, $k$, $V_1,\dots,V_n$ as
above,
is the space $E_\h(z,\tau,\lambda)$ of
linear functionals on $\Vhn$ annihilated by $\gz$.
If $\lambda=0$, then $E(z,\tau)=E_\h(z,\tau,0)$ is called
the space of {\em conformal blocks} at $(z,\tau)$.
\end{definition}

\noindent Let us vary the parameters: let $U$ be an open
subset of $\co\times\h$. Then the space of holomorphic
functions $\omega:U\to \Vhn{}^*$,
(i.e., of functions $\omega$ whose evaluation
$\langle\omega,u\rangle$ on any fixed vector $u\in\Vhn$ is
holomorphic on $U$), is a right $\Lie_\h(U)$-module.

\begin{definition}
The space $E_\h(U)$ of {\em holomorphic twisted conformal blocks
on $U\subset \co\times\h$} is the space of $\Vhn{}^*$-valued holomorphic
functions $\omega$, so that for all open subsets $U'$ of
$U$, the
restriction of $\omega$ to $U'$ is annihilated
by $\Lie_\h(U')$.
We also define  the space $E(U)$ of {\em holomorphic
 conformal blocks on $U\subset \co$} by replacing
$\Lie_\h$ by $\Lie$.
\end{definition}

\noindent
With this definition, the assignments $U\mapsto E_\h(U)$, $U\mapsto
E(U)$ are sheaves of $\cal O$-modules.

\begin{lemma}
Let $U$ be an open subset of $\co\times\h$ (resp.\  of $\co$).
Then $\omega\in E_\h(U)$ (resp.\ $E(U)$)
if and only if $\omega$ is holomorphic on $U$ and $\omega(x)\in E_\h(x)$
(resp.\ $E(x)$) for all $x\in U$.
\end{lemma}

\begin{proof}
It is obvious that if $\omega$ is holomorphic and if $\omega(x)\in E_\h(x)$
for all $x\in U$, then $\omega\in E_\h(U)$.
Let $\omega\in E_\h(U)$, and $x\in U$.
To show that $\omega(x)\in E_\h(x)$, we have to show that every
element $X$ of $\Lie(x)$ is the restriction of an element of
$\Lie_\h(U')$ for some neighborhood $U'$ of $x$.
But this follows from Prop.\ \ref{free}. The same applies
in the untwisted case.
\end{proof}\par\medskip

\section{Flat connections, theta functions}
\parag{The flat connection}
\label{connection}
For each open subset $U$ of $\co\times\h$ we have defined
a Lie algebra $\Lie_\h(U)$ acting on $\Vhn(U)$, the space
of holomorphic functions on $U$ with values in $\Vhn$.
It is convenient to extend this definition. Let
$G$ be the simply connected complex Lie group
whose Lie algebra is $\g$, and for $(z,\tau,g)\in\co\times G$,
let $\Lie(z,\tau,g)$  be
the Lie algebra of meromorphic $\g$-valued functions
$X(t)$, on the complex plane
whose poles modulo $L_\tau$
belong to $\{z_1,\dots,z_n\}$, and with
multipliers
\begin{gather*}
X(t+1)=X(t),\\
X(t+\tau)=\Ad(g)X(t).
\end{gather*}
If $U$ is an open subset of $\co\times G$, define
$\Lie_G(U)$ to be the Lie algebra of meromorphic functions
on $U\times\C\ni(z,\tau,g,t)$ whose poles are on the hyperplanes
$t=z_i+n+m\tau$, $n,m\in\Z$, and restricting
to functions in $\Lie(z,\tau,g)$ for fixed $(z,\tau,g)\in
U$. As above, we introduce the space $E_G(z,\tau,g)$ of
$\Lie(z,\tau,g)$-invariant linear forms on $\Vhn$, and
the sheaf $U\to E_G(U)$ of $\Lie_G(U)$ invariant holomorphic
$\Vhn{}^*$-valued functions.

Let $\eta(z,\tau,t)$ be a meromorphic function on $\co\times\C$
whose poles belong to the hyperplanes $t=z_i+n+m\tau$ and
such that, as function of $t\in\C$,
\begin{equation}\label{RR}
\eta(z,\tau,t+1)=\eta(z,\tau,t),
\qquad \eta(z,\tau,t+\tau)=\eta(z,\tau,t)-2\pi i.
\end{equation}
Although the construction does not depend on which $\eta$ we chose, we
will always set
\begin{gather*}
\eta(z,\tau,t)=\rho(t-z_1,\tau),\\
\rho(t,\tau)=\frac\partial{\partial t}\log \theta_1(t|\tau),\\
\theta_1(t|\tau)=-\sum_{j=-\infty}^{\infty}e^{\pi i(j+\half)^2\tau
+2\pi i(j+\half)(t+\half)},
\end{gather*}
for definiteness.

Let $A_Y(z,\tau,g,t)$ be a meromorphic function on $\co\times G\times
\C$, depending linearly on $Y\in\g$, whose poles as a function
of $t$ belong to $\{z_1,\dots,z_n\}$, and such that
\begin{gather}\label{AA}
A_Y(z,\tau,g,t+1)=A_Y(z,\tau,g,t),\\
A_Y(z,\tau,g,t+\tau)=\Ad(g)(A_Y(z,\tau,g,t)-Y).\notag
\end{gather}
If $\psi(A)=\sum_{j\in\Z}e^{i\pi(j+\half)^2\tau}(-A)^j$,
we may take $A_Y$ to be
\begin{equation}\label{AY}
A_Y(z,\tau,g,t)=
\frac1{1-\Ad(g^{-1})}
(1-\frac{\psi(e^{2\pi i(t-z_1)}\Ad(g^{-1}))}
{\psi(e^{2\pi i(t-z_1)})})Y.
\end{equation}
Note that for fixed $z,\tau,t$ and $Y$, $A_Y$ extends to a regular
function of $g\in G$.

Denote by $\partial_Y$ the derivative in the direction
of the left invariant vector field on $G$ associated to
$Y\in\g$: $\partial_Yf(g)=\lim_{\epsilon\to 0}\frac d{d\epsilon}
f(g\exp \epsilon Y)$, and by $\partial_{z_i}$, $\partial_\tau$,
$\partial_t$ the partial derivatives with respect
to the coordinates $z_i,\tau, t$ of $\co\times\C$.

The properties \Ref{RR}, \Ref{AA} imply
the

\begin{proposition}
Let $U$ be an open  subset of $\co\times G$. The
differential operators
\begin{gather*}
D_{z_i}X(x,t)=\partial_{z_i}X(x,t),\\
D_\tau X(x,t)=\partial_\tau X(x,t) -\frac1{2\pi i}\eta(x,t)
\partial_tX(x,t),\\
D_YX(x,t)=\partial_YX(x,t)+[A_Y(x,t),X(x,t)],
\qquad x=(z,\tau,g)\in U\times\C,
\end{gather*}
map $\Lie_G(U)$ to itself.
\end{proposition}
\noindent
Therefore, we have a connection $D:\Lie_G(U)\to\Omega^1(U)\otimes
\Lie_G(U)$,
defined by $D=\sum dz_j\otimes D_{z_j}+d\tau\otimes D_\tau
+\sum\theta_a\otimes D_{\theta^a}$, for any basis of left invariant
vector fields $\theta^a$ on $G$ with dual basis $\theta_a$.

We proceed to define a connection on $E_G$. Consider
first the following differential operators on the space $\Vhn(U)$ of
$\Vhn$-valued holomorphic functions on the open set $U\subset \co
\times G$.
\begin{gather*}
\label{cz}\nabla_{z_i}v(x)=\partial_{z_i}v(x)+L_{-1}^{(i)}v(x),\\
\label{ctau}\nabla_\tau v(x)=\partial_\tau v(x) -\frac1{2\pi i}\delta(
\eta(x)\partial_t)v(x),\\
\nabla_Yv(x)=\partial_Yv(x)+\delta(A_Y(x))v(x),\qquad x\in U.
\end{gather*}
In this formula the definition of the operator $\delta$ taking the
Laurent expansion at the points $z_i$ (see \Ref{PI}) is extended to
general meromorphic $\g$-valued functions and vector fields considered
as a function of the variable $t\in\C$. For a meromorphic vector field
$\xi=\xi(t)\frac d{dt}$ on the complex plane we set $\delta(\xi)=\Sigma
\delta_i(\xi)$, with $\delta_i(\xi)=\xi(z_i+t)\frac d{dt}
\in\C((t))\frac d{dt}$.
Let  $\nabla:\Vhn(U)\to\Omega^1(U)\otimes
\Vhn(U)$ denote the connection $\sum dz_j\otimes \nabla_{z_j}+
d\tau \otimes\nabla_\tau
+\sum\theta_a\otimes \nabla_{\theta^a}$,

\begin{proposition}
The connections $D$, $\nabla$ obey the compatibility condition
\be
\nabla(Xv)=(DX)v+X\nabla v,\qquad X\in\Lie_G(U),\quad v\in\Vhn.
\ee
\end{proposition}

\begin{proof} This is verified by explicit calculation.
\end{proof}\par\medskip

This has the following consequence. Define $\nabla$ on
holomorphic functions $\omega$ on $U$ with values in the dual
${\Vhn}^*$ (i.e., such that $\langle\omega(x),v\rangle$
is holomorphic on $U$ for all $v\in\Vhn$) by the formula
$\langle \nabla\omega(x),v(x)\rangle=d\langle\omega(x),v(x)\rangle
-\langle\omega(x),\nabla v(x)\rangle$.

\begin{corollary}
The connection
$\nabla$ preserves twisted holomorphic conformal blocks,
i.e., it maps $E_G(U)$ to $\Omega^1(U)\otimes E_G(U)$.
\end{corollary}

\begin{proposition}\label{flat}
The connection $\nabla$ on $E_G(U)$ is flat.
\end{proposition}

\begin{proof}
For $X,Y\in\g$, the curvature  $F(X,Y)=[\nabla_X,\nabla_Y]
-\nabla_{[X,Y]}$ is given by the expression
\be
F(X,Y)=\partial_X\delta(A_Y)-\partial_Y\delta(A_X)
+[\delta(A_X),\delta(A_Y)]-\delta(A_{[X,Y]}).
\ee
Note that the cocycle
\be
\int_\gamma (\frac d{dt}A_X,A_Y)dt,
\ee
vanishes: indeed, the integrand $I(t)$ is $\Z$-periodic
and obeys $I(t+\tau)=I(t)+\frac d{dt}g(t)$ for some
$\Z$-periodic function $g(t)$ and the integration
cycle $\gamma$ can be decomposed into a sum of contours bounding
some fundamental domains. The contributions of the four
edges cancel by periodicity, except for a term
$\int_x^{x+1}g'(t)dt=0$.

Thus we can write $F$ as
\begin{gather*}
F(X,Y)=\delta(\tilde F(X,Y)),\\
\tilde F(X,Y)=\partial_XA_Y-\partial_YA_X
+[A_X,A_Y]-A_{[X,Y]}.
\end{gather*}
Now, as a simple calculation shows, $\tilde F(X,Y)$, viewed
as a function of $t\in\C$ with values is $\g$, is $\Z$-periodic,
and obeys $\tilde F(X,Y)(t+\tau)
=\Ad(g)\tilde F(X,Y)(t)$, as a consequence
of \Ref{AA}. Thus $F(X,Y)$ is in the image of $\Lie_G(U)$,
and vanishes on invariant linear forms.

A similar reasoning applies to the commutators
$[\nabla_{z_i},\nabla_X]$, $[\nabla_\tau,\nabla_X]$,
$X\in\g$. These commutators are also in the image
of $\Lie_G(U)$ and thus vanish on invariant forms.
We are left with the commutator $[\nabla_\tau,\nabla_{z_i}]$,
which vanishes except possibly for $i=1$.
The proof that it  vanishes also for $i=1$ will be given later on
(see \ref{kzbe}).
\end{proof}\par\medskip

The group $G$ acts on $\Vhn$, since
the cocycle vanishes on $\g\subset\Lg$.
Denote this action simply $G\times\Vhn\ni (h,v)\mapsto hv$.

\begin{proposition}\label{conj}
For all $h\in G$, $X\to\Ad(h)X$ is a Lie algebra isomorphism from
$\Lie(z,\tau,g)$ to $\Lie(z,\tau,hgh^{-1})$. Thus the map
$X\mapsto\phi_hX$ with $\phi_hX(z,\tau,g)=\Ad(h)X(z,\tau,h^{-1}gh)$ is
an isomorphism from $\Lie_G(U)$ to $\Lie_G(U')$ for any open
$U\subset\co\times G$, where $U'=\{(z,\tau,hgh^{-1})| (z,\tau,g)\in
U\}$.  Moreover, for any $X\in\Lie_G(U)$, $\delta(\Ad(h)X)=h
\delta(X)h^{-1}$,
and thus $\rho_h\omega(z,\tau,g)=\omega(z,\tau,hgh^{-1})h$ defines an
isomorphism $\rho_h:E_G(U')\to E_G(U)$. This isomorphism maps
horizontal sections to horizontal sections.
\end{proposition}

\begin{proof}
The first statement follows immediately from the definitions.
The fact that  $\delta(\Ad(h)X)=h\delta(X)h^{-1}$  is also
clear, once one notices that the 2-cocycle defining
the central extension vanishes if one of the arguments
is a constant Lie algebra element.
Finally $h$ commutes with $\nabla_{z_i}$ and
$\nabla_\tau$, and we have $\nabla_X\rho_h=
\rho_h\nabla_{\Ad(h)X}$, $X\in\g$. The latter
identity follows from the equality (see \Ref{AY})
\be
\Ad(h)A_X(z,\tau,g,t)=
A_{\Ad(h)X}(z,\tau,hgh^{-1},t).
\ee
Thus $\rho_h$ preserves horizontality.
\end{proof}\par\medskip

The existence of a  connection implies, as in
\cite{TUY}, that
the sheaf $U\mapsto E_G(U)$ is (the sheaf of holomorphic
sections of) a holomorphic vector bundle whose
fiber over $x$ is $E_G(x)$. This
follows once one notices that $E_G(U)$ is actually
a subsheaf of a locally free finitely generated sheaf
carrying a connection whose restriction to $E_G$ is $\nabla$.
Details on this point are in Appendix \ref{B}.

To make connection with the previous sections, consider
the pull back of $E_G$ by the map $\lambda\mapsto
\exp(2\pi i\lambda)$, from $\h$ to $G$. It is the
vector bundle $E_\h$ on $\co\times\h$. Let us
introduce coordinates $\lambda_\nu$ on $\h$ with respect to some
orthonormal basis ($h_\nu$). Then the pull-back
connection on $E_\h$ is given by \Ref{cz}, \Ref{ctau},
and, in the  direction of $\lambda$,
\be
\nabla_{\lambda_\nu}=\partial_{\lambda_\nu}
-\delta(h_\nu\rho(\cdot-z_1,\tau)).
\ee
Moreover, we can use the connection
 to identify by parallel translation the
space of conformal blocks $E(U)$ with the
space of twisted conformal blocks $\omega$ in $E_G(U\times G)$
(or in $E_\h(U\times\h)$)
such that $\nabla_X\omega=0$, $X\in\g$
 (or $\nabla_{\lambda_\nu}\omega=0$,
respectively).  Here we use the fact that $G$ and $\h$ are
simply connected.

The point of this construction is given by the
following result. Let $\Vn{}^*(U)$ be the space
of holomorphic functions on an open set $U$ with
values in the finite dimensional space $\Vn{}^*$.
We also set, for any open subset $U$ of $\co\times G$,
or $\co\times\h$, respectively,
\begin{gather*}
E_G(U)^\hor=\{\omega\in E_G(U)|\nabla_X\omega=0,\quad
\forall X\in\g\},\\
E_\h(U)^\hor=\{\omega\in E_\h(U)|\nabla_X\omega=0,\quad
\forall X\in\h\}.
\end{gather*}

\begin{proposition}\label{iota}
The compositions
\begin{gather*}
\iota_G:E(U)\to E_G(U\times G)^\hor\to\Vn{}^*(U\times G),\\
\iota_\h:E(U)\to E_\h(U\times\h)^\hor\to\Vn{}^*(U\times\h),
\end{gather*}
where the first map sends a
holomorphic conformal block  $\omega$ to the unique
twisted holomorphic conformal block horizontal
in the $G$ (resp.\ $\h$) direction, which coincides
with $\omega$ on $U\times\{1\}$ (resp.\ $U\times\{0\}$), and
the second map is the restriction to $\Vn$,
are injective.
\end{proposition}

\begin{proof} The first map is an isomorphism to
the space of twisted holomorphic conformal blocks horizontal
in the $G$ (resp.\ $\h$) direction. The fact that
the second map
is injective follows from the fact that using the
invariance and the equation $\nabla_X\omega=0$
(resp.\ $\nabla_{\lambda_\nu}\omega=0$), one can express
$\langle\omega,v\rangle$ for any $v\in\Vhn$
linearly in terms of the restriction of $\omega$
to $\Vn$.\end{proof}\par\medskip

We may (and will) thus view the sheaf $E$ of sections
of the vector bundle on $\co$ of conformal blocks  as a subsheaf
of $\Vn(U\times \h)$. The next steps are a characterization
of this subsheaf and a formula
for the connection after this identification.

\parag{Theta functions}
Let $Q^\vee=\{q\in\h|\exp(2\pi iq)=1\in G\}$ be the coroot
lattice of $\g$.

\begin{definition} Let $(z,\tau)\in\co$, and $V_1$,\dots,
$V_n$ be finite dimensional $\g$-modules, and $k$ a
non-negative integer. The
space $\Theta_k(z,\lambda)$  of theta functions of level
 $k$ is the space of
holomorphic functions $f:\h\to\Vn{}^*$ such that
\begin{enumerate}
\item[(i)] $\sum_{i=1}^nf(\lambda)h^{(i)}=0$.
\item[(ii)] One has the following transformation properties
unter the lattice $Q^\vee+\tau Q^\vee\subset\h$:
\begin{eqnarray*}
f(\lambda+q)&=&f(\lambda)\\
f(\lambda+q\tau)&=&f(\lambda)
\exp\bigl(-\pi ik(q,q)\tau-2\pi ik(q,\lambda)-2\pi i\sum_{j=1}^n
z_jq^{(j)}\bigr)
\end{eqnarray*}
\end{enumerate}
\end{definition}

\noindent
The space of such theta functions is finite dimensional,
as can be easily seen by Fourier series theory.
 Denote by $W$ be the Weyl group of $\g$, generated
by reflection with respect to root hyperplanes.
It is known that this group is isomorphic to $N(H)/H$,
$N(H)\subset G$ being the normalizer of $H=\exp(\h)$.
For $w\in W$, let $\hat w\in N(H)$ be any representative of the
class of $w$
in $N(H)/H$.  The Weyl group acts on the space of theta
functions. Indeed, if $f\in\Theta_k(z,\tau)$, then
$(wf)(\lambda)=f(w^{-1}\lambda)\hat w^{-1}$ also
in  $\Theta_k(z,\tau)$, (the coroot lattice and the invariant
bilinear form are preserved by the Weyl group), and
is independent of the choice of representative $\hat w$ by
(i).
Let $\Theta_k(z,\tau)^W$ denote the space of $W$-invariant
theta functions.

\begin{thm}\label{main} Let $\g=A_l$, $l\geq 2$,
$D_l$, $l\geq 4$,
$E_6$, $E_7$, $E_8$, $F_4$, or $G_2$.
Then the image of $\iota_\h$ is contained in
the space of holomorphic functions $\omega\in\Vn{}^*(U\times\h)$
such that for all $(z,\tau)\in\co$, $\omega(z,\tau,\cdot)$
belongs to $\Theta_k(z,\tau)^W$, and such that for all roots $\alpha$,
$X\in\g_\alpha$ and nonnegative integers $p$,
\be
\omega(z,\tau,\lambda)X^p=O(\alpha(\lambda)^p),
\ee
as $\alpha(\lambda)\to 0$.
\end{thm}

\noindent In the remaining cases, we have

\begin{thm}\label{main1} Let $\g=A_1$, $B_l$ or $C_l$, $l\geq 2$.
Then the image of $\iota_\h$ is contained in
the space of holomorphic functions $\omega\in\Vn{}^*(U\times\h)$
such that for all $(z,\tau)\in\co$, $\omega(z,\tau,\cdot)$
belongs to $\Theta_k(z,\tau)^W$, and such that for all $\alpha\in\Delta$,
$r,s\in\{0,1\}$,
$X\in\g_\alpha$ and nonnegative integers $p$,
\be
\omega(z,\tau,\lambda)
\exp\left(2\pi i\,c_{r,s}\sum_jz_j\lambda^{(j)}\right)
X^p=O((\alpha(\lambda)-r-s\tau)^p),
\ee
as $\alpha(\lambda)\to r+s\tau$, with  $c_{r,0}=0$, $c_{r,1}=(r+\tau)^{-1}$.
\end{thm}

\noindent The proof of these theorems will be completed in \ref{prt}.
We conjecture that the space of functions described in
Theorems \ref{main}, \ref{main1} actually coincides with the image
of $\iota_\h$. This conjecture is verified in a simple
class of examples in \ref{example} below.

The fact that the formulation of the result is simpler
for certain Lie algebras is due to the following
property shared by the Lie algebras of Theorem \ref{main}:
for each root $\alpha$ and integer $m$ there exist
an element $q$ in the coroot lattice with $\alpha(q)=m$.
For the other simple Lie algebras this is true only if
$m$ is assumed to be even. More on this in \ref{vanishing}.

\parag{Affine Weyl group}
 Since $H$ acts trivially on $E_\h$, the
Weyl group acts (on the right)  on the values of $E_\h$: $w$ acts
as $\hat w$, and this is independent of the choice of
representative.

\begin{proposition}\label{AWG}
Let $\omega\in E_\h(U\times\h)^\hor$. Then, for all
$q\in Q^\vee$,
\be
\omega(z,\tau,\lambda+q)=\omega(z,\tau,\lambda).
\ee
For all $w\in W$,
\be
\omega(z,\tau,w\lambda)=\omega(z,\tau,\lambda)\hat w^{-1}.
\ee
\end{proposition}

\begin{proof}
The value of $\omega$ at $(z,\tau,\lambda+q)$ is obtained
from $\omega(z,\tau,\lambda)$ by parallel transport along
some path from $\lambda$ to $\lambda+q$.
Recall that $\omega$ is the pull back of a section of
$E(U\times G)^\hor$ to $U\times \h$. The image of the
path in $G$ is closed, and contractible ($G$ is simply connected),
which proves the first claim.

{}From Prop.\ \ref{conj} and the fact that $w\cdot\lambda=\Ad(\hat w)\lambda$,
we see that if $\omega$ is horizontal
then also $\rho_{\hat w}\omega$ is horizontal. But these
horizontal sections coincide at $\lambda=0$, and thus everywhere.
\end{proof}\par\medskip

\parag{Modular transformations}
The group $\SL(2,\Z)$ acts as follows on $\co\times\h$:
if $A=\left(\begin{matrix}a&b\\ c&d\end{matrix}\right)
\in \SL(2,\Z)$,
\be
A\cdot(z,\tau,\lambda)=
\left(\frac z{c\tau+d},\frac{a\tau+b}{c\tau+d},
\frac\lambda{c\tau+d}\right).
\ee
\begin{lemma} Introduce the linear functions
 $\ell_\lambda(t)=2\pi i\lambda t$ for $\lambda\in\h$.
For all $x\in\co\times \h$, $A\in \SL(2,\Z)$,
the map $X\to\phi_A(x)X$ with
\be
(\phi_A(x)X)(t)=\exp(-\ad c\, \ell_\lambda(t))X((c\tau+d)t),\qquad
A=\left(\begin{matrix}a&b\\ c&d\end{matrix}\right),
\ee
is a Lie algebra isomorphism from $\Lie(x)$ to
$\Lie(Ax)$.
\end{lemma}

\begin{proof} This follows directly from the definitions.
\end{proof}\par\medskip

We have defined an action $\Lie(x)\otimes\Vhn\to\Vhn$, for
all $x\in\co\times\h$. Let us denote it as $X\otimes v
\mapsto \delta_x(X)v$ (see \Ref{PI}) to emphasize the
$x$-dependence.

\begin{lemma}\label{inter}
Define a linear isomorphism $v\to\rho_A(x)v$ from $\Vhn$ viewed
as  $\Lie(x)$-module to $\Vhn$ viewed as  $\Lie(Ax)$-%
module: if $x=(z,\tau,\lambda)$,
\be
\rho_A(x)=\eta_A(x)(c\tau+d)^{-\sum_{j=1}^nL_0^{(j)}}
\exp\left(-\frac {c\,\delta_x(\ell_\lambda)}{c\tau+d}\right),
\qquad \eta_A(x)=
\exp\left(-{\frac{\pi i ck(\lambda,\lambda)}{c\tau+d}}\right).
\ee
This map has the intertwining property
\be
\rho_A(x)\delta_x(X)=\delta_{Ax}(\phi_A(x)X)\rho_A(x),
\ee
for all $X\in\Lie(x)$.
\end{lemma}

\noindent Note that the choice of coefficient $\eta_A$ is
irrelevant for the validity of the Lemma. However, it
is important for compatibility with the connection,
see below.

We should also add a remark about the power of $(c\tau+d)$.
The exponent $\Sigma L_0^{(i)}$ is diagonalizable with
finite dimensional eigenspaces. However the eigenvalues are
frectional in general, and the power is defined for a choice
of branch of the logarithm for each $A\in\SL(2,\Z)$. This
is made more systematic in the next subsection.

\begin{proof}
This is again straightforward. The only subtlety is that,
a priori, there could be a contribution from the central
extension in the computation of the intertwining property.
However the central term appearing in this computation
is proportional to the sum of the residues of the
component of $\partial_tX$ along $\lambda$, which is doubly
periodic. By the residue theorem, this sum vanishes.
\end{proof}\par\medskip

\noindent We can thus define linear maps
$\omega\mapsto\rho_A^*\omega$
by
 $\rho_A^*\omega(x)=\omega(Ax)\rho_A(x)$. Lemma \ref{inter}
implies then that $\rho^*_A$
is an isomorphism from $E_\h(U)$ to $E_\h(A^{-1}(U))$.

\begin{lemma}\label{coco}
 Let $A^*$ be the pull back on one-forms of
the map $x\mapsto Ax$ defined on some open $U\subset\co\times\h$,
and $\nabla:E_\h(U)\to\Omega^1(U)\otimes E_\h(U)$ be the
connection defined in  \ref{connection}.
We have
\be
\nabla\rho^*_A=A^*\otimes\rho^*_A\nabla.
\ee
\end{lemma}

\noindent This fact can be derived from a straightforward but
unfortunately lengthy calculation. The main identity one
uses is
\be
\rho\left(\frac t{c\tau+d},\frac{a\tau+b}{c\tau+d}\right)
=(c\tau+d)\rho(t,\tau)+2\pi ict.
\ee
Lemma \ref{coco} ensures that $\rho^*_A$ maps horizontal
sections to horizontal sections. Moreover, since $A_*\partial_{\lambda_\nu}
=(c\tau+d)^{-1}\partial_{\lambda_\nu}$ does not have components
in $z$ or $\tau$ direction, $\rho_A^*$ maps sections which are horizontal in
the $\h$ direction to sections with the same property:
\be
\rho^*_A:E_\h(U\times\h)^\hor\to E_\h(A^{-1}U\times\h)^\hor.
\ee
Let us apply this in the special case
 $A=\left(\begin{matrix}0&1\\ -1&0\end{matrix}\right)$.

\begin{proposition}\label{TAU}
Let $\omega\in\Vn{}^*(U)$ be in the image of $\iota_\h$.
Then for all $q\in Q^\vee$,
\be
\omega(z,\tau,\lambda+\tau q)=
\omega(z,\tau,\lambda)\exp(-2\pi i(q,\lambda)k-\pi i(q,q)k\tau
-2\pi i\sum_{j=1}^nz_j q^{(j)}).
\ee
\end{proposition}
 \begin{proof}
We have $\rho^*_A\omega\in E_\h(A^{-1}U\times\h)^\hor$. Thus,
\be
\rho^*_A\omega(\frac z\tau,-\frac1\tau,\frac\lambda\tau+q)
=\rho^*_A\omega(\frac z\tau,-\frac1\tau,\frac\lambda\tau),
\ee
for all coroots $q$, by Lemma \ref{AWG}, and $(z,\tau)\in U$.
Explicitly,
\be
\omega(z,\tau,\lambda+q\tau)
\rho_A(\frac z\tau,-\frac1\tau,\frac\lambda\tau+q)
=
\omega(z,\tau,\lambda)
\rho_A(\frac z\tau,-\frac1\tau,\frac\lambda\tau).
\ee
Inserting the formula for $\rho_A$ we obtain
\be
\omega(z,\tau,\lambda+q\tau)\tau^{\Sigma_jL_0^{(j)}}
=\omega(z,\tau,\lambda)\tau^{\Sigma_jL_0^{(j)}}
e^{-2\pi i(q,\lambda)k-\pi i(q,q)k\tau}\exp(-\tau\delta_x(\ell_q)),
\ee
with $x=(z/\tau,-1/\tau,\lambda/\tau)$.
Now we use the fact that, on $\Vn$, $\delta_x(\ell_q)$ acts as
$\Sigma_j2\pi i( z_j/\tau)q^{(j)}$, and that $L_0^{(j)}$ acts
as a multiple of the identity, to conclude the proof.
\end{proof}\par\medskip

\parag{Monodromy (projective) representations of $\SL(2,\Z)$}
In this subsection, we assume that $n=1$, set $z_1=0$, and show that a
central extension of $\SL(2,\Z)$ acts on the space of horizontal
sections of the bundle of conformal blocks.

The fact that we have a central extension comes from the necessity to
choose a branch of the logarithm to define the expression
$(c\tau+d)^{L_0}$. In fact $L_0$ is diagonalizable with finite
dimensional eigenspaces, and any two eigenvalues differ by an
integer. Moreover, $L_0$ acts by a non-negative rational multiple
(Cas$(V)/(k+h^\vee$)) of the identity on $V\subset V^\wedge$ for any
integrable $V^\wedge$ of level $k$. Let $L_0|_V=\frac rs
\operatorname{id}_V$,
$r,s\in\N$. We introduce a central extension
\be
0\to\Z/s\Z\to\Gamma_s\to \SL(2,\Z)\to 1,
\ee
of $\SL(2,\Z)$ by the cyclic group of order $s$. The group
$\Gamma_s$ consists of pairs $(A,\phi)$ where $A\in SL(2,\Z)$
has matrix elements $a,b,c,d$ and $\phi$ is a holomorphic
function on the upper half plane such that $\phi(\tau)^s=
c\tau+d$. The product is
$(A,\phi)(B,\psi)=(AB,\phi\circ A\cdot\psi$).
Then this group acts on $V^*$ valued functions
on $\Hp\times\h$ as above, but keeping track of
the choice of branch:
\be
(A,\phi)^{-1}\omega(\lambda,\tau)=\omega(A\cdot(\lambda,\tau))
\eta_A(\lambda,\tau)\phi(\tau)^{-r}.
\ee
This action preserves the connection. (The inversion here is
to correct for the ``wrong'' order $\rho_A^*\rho_B^*=\rho_{BA}^*$,
up to ambiguity in the choice of branch). Thus we conclude
that
$\Gamma_s$ acts on the  space of global horizontal sections
on $\Hp\times\h$ of $E_\h$. This monodromy representation
restricts to the character $[m]\mapsto\exp(2\pi imr/s)$
of $\Z/s\Z$.

In the case of $V=$ trivial representation, this monodromy
representation is just the representation of $SL(2,\Z)$ on characters of
affine Lie algebras (see \cite{Ka}).  It would be interesting to
calculate this monodromy representation explicitly for general $V$. Some
progress in the $sl_2$ case was made in \cite{CrFeWi}, where a
connection with the adjoint representation of the corresponding quantum
group was established.

\parag{The vanishing condition}\label{vanishing}
Let $G$ be a simple complex Lie group with Lie
algebra $\g$, $\h$ a
Cartan subalgebra, $\g=\h\oplus\oplus_{\alpha\in\Delta}\g_\alpha$
a Cartan decomposition and $H=\exp\h$. Suppose that
$\rho:G\to\operatorname{End}(V)$ is a finite
dimensional representation of $G$.
Thus $V$ is also a $\g$-module.
For $K=G$ or $H$ let $I(K,V)$ be the space of holomorphic functions on
$K$ with values in $V$ such that $\forall g,h\in K$,
$u(ghg^{-1})=\rho(g)u(h)$. The Weyl group $W$ acts on $I(H,V)$:
let $\hat w$ be a representative of $w\in W$ in $N(H)$. Then
$(wf)(h)=\rho(\hat w)f(w^{-1}\cdot h)$ is well defined for
$f\in I(H,V)$, since $H$ acts trivially on the image of functions
in $I(H,V)$. We denote by $I(H,V)^W$ the space of Weyl-invariant
functions in $I(H,V)$.

\begin{lemma}
The restriction map $I(G,V)\to I(H,V)$ is injective. Its
image is the space $I_0(H,V)^W$
of functions $u$ in $I(H,V)^W$
such that for all positive roots $\alpha$, $X\in\g_\alpha$
$p\in\N$, and $m\in\Z$
\begin{equation}\label{van}
X^pu(\exp 2\pi i\lambda)=O((\alpha(\lambda)-m)^p),
\end{equation}
as $\alpha(\lambda)\to m$.
\end{lemma}

\begin{proof}
The behavior of functions in $I(G,V)$ under conjugation
by $N(H)$ implies Weyl invariance.

Let $X\in\g_\alpha$ and $\lambda\in\h$. Then
\begin{equation}\label{qwe}
\Ad(\exp( 2\pi i\lambda))\,X=e^{2\pi i\alpha(\lambda)}X.
\end{equation}
If $u\in I(G,V)$, $u(\exp(X)\exp(2\pi i\lambda))$
is a holomorphic $Q^\vee$-periodic function of $\lambda\in\h$,
(thus a holomorphic function on $H$). On the other
hand, by \Ref{qwe}
\bea
u(\exp(X)\exp(2\pi i\lambda))
&=&
u\left(\exp\left(\frac X{1-e^{2\pi i\alpha(\lambda)}}\right)
  \exp(2\pi i\lambda)
  \exp\left(-\frac X{1-e^{2\pi i\alpha(\lambda)}}\right)\right)\\
 &=&
\rho(\exp((1-e^{2\pi i\alpha(\lambda)})^{-1}X))
u(\exp(2\pi i\lambda))\\
&=&
\sum_{p=0}^M\frac1{p!}(1-e^{2\pi i\alpha(\lambda)})^{-p}X^p
u(\exp(2\pi i\lambda)),
\eea
for some $M$. We see that the latter expression is
holomorphic on the affine hyperplanes $\alpha(\lambda)=m$,
if and only if, for all $p$, $X^pu$ vanishes there to order
at least $p$.

To conclude the proof, we use some facts about conjugacy classes in
algebraic groups (see, e.g., \cite{St}, Chapter 3). Let, for each root
$\alpha$ and integer $m$, $H_{\alpha,m}\subset H$ be the set of elements
of the form $\exp(2\pi i\lambda)$ such that $\alpha(\lambda)=m$. The
conjugacy classes containing elements in $H_\reg=H-\cup H_{\alpha,m}$
form the dense open subset $G_\reg$ of regular semisimple
elements in $G$. Its complement contains the
set $H_1$ consisting of conjugacy classes of elements of the form
$\exp(X)\exp(2\pi i\lambda)$, where $\lambda$ lies on precisely one of
the distinct $H_{\alpha,m}$. This elements are regular, as they are
regular in the identity component of the stabilizer of $\exp(2\pi
i\lambda)$, (see \cite{St}, 3.5), which is the direct product of a torus
of dimension rank-1 times the $\SL(2)$ subgroup associated with
$\alpha$. By the above reasoning, a Weyl invariant function on $H$
extends uniquely to an equivariant holomorphic function on $G_\reg$, and
the vanishing conditions imply that it extends to a holomorphic
functions on $G_\reg\cup H_1$. The complement of $G_\reg\cup H_1$
consists of higher codimension classes whose closure intersects $H_1$,
and of classes whose closure do not intersect $H_\reg\cup H_1$. Counting
dimensions shows that this complement is of codimension at least two, so
by Hartogs' theorem, our vanishing conditions are sufficient to have an
extension to all of $G$.
\end{proof}

\medskip

 By Weyl invariance, we may replace
the set of positive roots in the formulation of the
lemma to a subset of roots consisting of one root
for each Weyl group orbit.
 Also, we may restrict the values of $m$, by
$Q^\vee$ periodicity of $u(\exp 2\pi i\lambda)$. Indeed,
if the vanishing condition holds at $\alpha(\lambda)=m$,
it also holds at $\alpha(\lambda)=m-\alpha(q)$ for all
$q\in Q^\vee$.

We thus have the following result. The action of
the affine Weyl group $W^\wedge=W\tilde\times Q^\vee$
on $\Delta\times\Z$ is defined by
\be
(w,q)(\alpha,m)=(w\alpha,m-\alpha(q)).
\ee

\begin{lemma}
The subspace $I_0(H,V)^W\subset I(H,V)^W$ is characterized
by the vanishing condition  \Ref{van}, for $(\alpha,m)$
in any fundamental domain for the action of $W^\wedge$
on $\Delta\times\Z$.
\end{lemma}

\noindent From \cite{bou}, we see that in the cases
$A_l$, $l\geq 2$,
$D_l$, $l\geq 4$,
$E_6$, $E_7$, $E_8$, $F_4$, $G_2$ a fundamental domain
is $\{(\alpha,0), \alpha\in F\}$, where $\alpha$ runs
over a fundamental domain $F$ (consisting of one
or two elements) of $W$.
If $\g=A_1$, $B_l$, $C_l$, then we have to add $(\alpha,1)$
where $\alpha$ is a long root.

As a corollary we obtain a more precise characterization of the
image of $i_\h$. Let us identify functions on $H$ with
$Q^\vee$-periodic functions on $\h$ via the map
$\lambda\mapsto \exp(2\pi i\lambda)$, and view $V^{[n]*}$
as a representation of $G$ by $\langle\rho(g)u,v\rangle
=\langle u,g^{-1}v\rangle$.

\begin{corollary}\label{lastbutone}
 The image of $E(U)$ by $\iota_\h$ is contained in
the space of functions $\omega\in V^{[n]*}(U\times\h)$
such that for all $(z,\tau)\in C^{[n]}$,
$\omega(z,\tau,\cdot)$ belongs to $I_0(H,V^{[n]*})^W$.
\end{corollary}

\noindent Moreover, if $\omega\in E_\h(U\times\h)^\hor$, then
$\rho^*_A\omega\in E_\h(A^{-1}U\times\h)^\hor$, implying
further vanishing conditions: let $x=(z,\tau,\lambda)$
and $A=\left(\begin{matrix}a&b\\ c&d\end{matrix}\right)$.
On $\Vn$ $L_0^{(j)}$ acts by a scalar $\Delta_j$.
 The restriction of $\rho_A^*\omega$ to $\Vn$ is
\begin{displaymath}
(\rho^*_A\omega)(x)=
\omega(Ax)(c\tau+d)^{-\Sigma_j \Delta_j}
\eta_A(x)
e^{-\frac c{c\tau+d}\sum_jz_j\lambda^{(j)}}.
\end{displaymath}
It follows that, for all $p$,
\begin{displaymath}
\omega(Ax)\exp({-\frac c{c\tau+d}\sum_jz_j\lambda^{(j)}})
X^p=O((\alpha(\lambda)-m)^p),
\end{displaymath}
if $\alpha(\lambda)\to m$. Changing variables, this
implies that
\begin{displaymath}
\omega(z,\tau,\lambda)
\exp\left(-2\pi i \frac c{a-c\tau}\sum_jz_j\lambda^{(j)}\right)
X^p=O((\alpha(\lambda)-m(a-c\tau))^p).
\end{displaymath}
Since any pair $a$, $c$ of relatively prime integers appear in
the first column of some SL(2,$\Z$) matrix, we obtain the result

\begin{corollary}\label{last} The image of $E(U)$ by
$\iota_\h$ is contained in
the space of functions $\omega\in V^{[n]*}(U\times\h)$ such that for all
$(z,\tau,\lambda)\in C^{[n]}$, $r$, $s$, $p\in\Z$, $p\geq 1$, $(r,s)\neq
(0,0)$,
\begin{equation}\label{vaco}
\omega(z,\tau,\lambda)
\exp(2\pi i\frac s{r+s\tau}\sum_jz_j\lambda^{(j)})
X^p=O((\alpha(\lambda)-r-s\tau)^p),
\end{equation}
as $\alpha(\lambda)\to r+s\tau$.
\end{corollary}

\parag{Proof of Theorems \ref{main}, \ref{main1}}\label{prt}
Theorems \ref{main}, \ref{main1} follow from Propositions \ref{AWG},
\ref{TAU},  and Corollaries \ref{lastbutone}, \ref{last}
together with the fact that twisted conformal
blocks are annihilated by $\h\subset\Lie_\h(U)$.

\parag{Examples}\label{example}
Here we give an explicit description of the space of
conformal blocks in some special cases. The discussion
parallels the constructions in \cite{FaGa}, where
Chern--Simons states in the case of $sl_2$ are
studied.
First of all consider the case of one point $z_1$ with
the trivial representation. Then the vanishing condition
is vacuous, and we are left to classify scalar Weyl
invariant theta functions of level $k$. This space
coincides with the space spanned by characters of
irreducible highest weight $\Lg$-modules, in accordance
with the Verlinde formula.

Next, we consider
the case of one point $z_1$, with a symmetric tensor
power of the defining  representation  $\C^N$ of $sl_N$.

If $N\geq 3$, the problem is reduced to describing the space of
Weyl invariant theta functions $\omega$ of level $k$, with
the property that
\be
e_\alpha^pu(\alpha(\lambda))=O(\alpha(\lambda)^p), \quad
\alpha(\lambda)\to 0,
\ee
for all $p=1,2,\dots$ and root vectors $e_\alpha\in\g_\alpha$.
 Actually
it is sufficient to consider one root $\alpha$, since the
Weyl group acts transitively on the set of roots of $sl_N$.

The symmetric power $S^j\C^N$ has a
non-zero weight space if and only if $j$ is a multiple
of $N$. Let us set $j=lN$, and denote by $\epsilon_i$ the
elements of the standard basis of $\C^N$. Then the weight zero subspace
of $S^{lN}\C^N=(\C^N)^{\otimes l}/S_N$ is one-dimensional and is spanned
by the class of $v=\epsilon_1^{\otimes l}\otimes
\cdots\otimes \epsilon_N^{\otimes l}$. The following considerations
apply also to the case $l=0$, if we agree that $S^0\C^N$ is
the trivial representation.

The Weyl group of $sl_N$ is the symmetric group $S_N$ and
is generated by adjacent transpositions $s_j$, $j=1,\dots, N-1$.
If we identify the Weyl group with $N(H)/H$, then a representative
in $N(H)$ of $s_j$ is given by $\hat s_j\epsilon_r=\epsilon_r$, if
$r\neq j,j+1$, $\hat s_j\epsilon_j=\epsilon_{j+1}$,
$\hat s_j\epsilon_{j+1}=-\epsilon_j$. It follows that $S_N$ acts
on the weight zero space by the $l$th power of the
alternating representation:
$\hat w v={{\epsilon}}(w)^l v$.

The next remark is that $e_\alpha^{l+1} v=0$ but $e_\alpha^lv
\neq 0$. We thus see that $\omega(\alpha(\lambda))
=O(\alpha(\lambda)^{l})$ as $\lambda$ approaches the
hyperplane $\alpha(\lambda)=0$. If $\omega$ is a Weyl-invariant
 theta function,
it then also vanishes to order $l$ on all hyperplanes $\alpha(\lambda)=
n+m\tau$, $n$, $m\in\Z$. Therefore the quotient of $\omega$  by
the $l$th power of the Weyl--Kac denominator $\Pi(\lambda,\tau)$
is an entire function, as $\Pi$ has simple zeroes on those
hyperplanes. Moreover, $\Pi$ is a (scalar) theta function of level
$N$ (the dual Coxeter number of $sl_N$), and $\Pi(w\lambda,\tau)=
{{\epsilon}}(w)\Pi(\lambda,\tau)$.

We conclude that the space of conformal blocks at fixed $\tau$
is contained
in the space of functions of the form
\begin{equation}\label{opu}
\omega(\lambda)=\Pi(\lambda,\tau)^lu(\lambda) \,v,
\end{equation}
where $u$ is an entire $Q^\vee$-periodic scalar function on $\h$,
such that $u(w\lambda)=u(\lambda)$, for all $w\in S_N$ and
\begin{eqnarray}
u(\lambda+q\tau)&=&\alpha(q,\lambda,\tau)^{k-Nl}u(\lambda),\label{qp} \\
\alpha(q,\lambda,\tau)&=&
\exp(-2\pi i(q,\lambda)-\pi i(q,q)\tau),\qquad q\in Q^\vee.\notag
\end{eqnarray}
We have assumed here that $N\geq 3$. In the $\sl_2$ case, where
the vanishing condition must be satisfied also at 3 other
points on $\h$, one can proceed in the same way, noticing that
the Weyl denominator vanishes there too.

A basis of $Q^\vee$-periodic functions with multipliers \Ref{qp}
is easily given using Fourier series. The basis elements $\theta_\mu$
are labeled by $\mu\in P/(k-Nl)Q^\vee$, where the weight lattice
$P$ is dual to $Q^\vee$ (if $k<N$ there are no
non-zero conformal blocks). The Weyl group acts as
$\theta_\mu(w^{-1}\lambda)=\theta_{w\mu}(\lambda)$.

Therefore the dimension of our space is the number of
orbits of the Weyl group in $P/(k-Nl)Q^\vee$. This number
is well-known: a fundamental domain in $P$ for the action of
the semidirect product of the Weyl group by the group
of translations by $(k-Nl)Q^\vee$ is the set of
weights in the (dilated) Weyl alc\^ove
$I_{k-Nl}$, see \Ref{ik}.

 More explicitly, if $\alpha_i$ are simple roots,
$\omega_i$ fundamental
weights with $(\omega_i,\alpha_j)=\delta_{ij}$, and
$\mu=\Sigma_i n_i\omega_i$, then $\mu\in I_{k-Nl}$
if and only if the integers $n_i$ satisfy the inequalities
\be
n_i\geq 0,\quad i=1,\dots,N-1,\qquad \sum_{i=1}^{N-1}n_i
\leq k-Nl.
\ee
The number of $N-1$-tuples of integers with these properties
is calculated to be
\be
\left(\begin{matrix}
k-N(l-1)-1\\ N-1
\end{matrix}\right).
\ee
This is the formula for the dimension of the space of
Weyl-invariant theta functions of level $k$ extending
to holomorphic functions on SL${}_N$, with values
in the $(l\cdot N)$th
symmetric power of the defining representation
of $sl_N$.
We now show\footnote{We learned how to do this computation
from H. Wenzl} that this coincides with the Verlinde formula \cite{Ver},
which according to \cite{TUY}, \cite{Fa} give the dimension
of the space of conformal blocks.

Let $I_k$ be the set of integrable highest weights of
level $k$. It consists of dominant integral weights
$\mu$ with $(\mu,\theta)\leq k$.
The dimension of the space of conformal blocks with one
point, to which an irreducible
 representation of highest weight $\mu\in  I_k$ is attached,
is given by the formula
\be
d_\mu=\sum_{\nu\in I_k}N^\nu_{\nu\mu},
\ee
in terms of the structure constants $N^a_{bc}$ of
Verlinde's fusion ring. A convenient formula for these constants
in terms of the classical fusion coefficients $\No^a_{bc}$
($=$ the multiplicity of $a$ in the decomposition of the
tensor product of $b$ with $c$) was given in \cite{GW},
and \cite {Ka}, Exercise 13.35.

Let $W_k^\wedge\simeq W^\wedge$
 be the group of affine transformations of
$\h^*$ generated by the Weyl group $W$ and the reflection
$s_0$ at the hyperplane $\{\lambda\in\h^*|(\theta,\lambda)=k+h^\vee\}$
($\theta$ is the highest root and $h^\vee$ the dual
Coxeter number). Let $\rho$ be half the sum of the positive
roots of $\g$ and define another action of $W_k^\wedge$ on
$\h^*$ by $w*\lambda=w(\lambda+\rho)-\rho$. Let
$\epsilon: W_k^\wedge\to\{1,-1\}$ be the homomorphism
taking reflections to $-1$.

 Then, for all $a,b,c\in I_k$,
\begin{equation}\label{sss}
N^a_{bc}=\sum_{w\in W_k^\wedge}\No^{w*a}_{bc}.
\end{equation}
Actually, in Verlinde's formula the coefficients
\newcommand{\s}[2]{\frac{S_{#1#2}}{S_{#1 0}}}
 $N^a_{bc}$ are given in terms of
 modular transformation properties
of characters. They are uniquely determined by the equation
\begin{displaymath}
\s db\,\s dc=\sum_a N^a_{bc}\,\s da,
\end{displaymath}
where, according to \cite{Ka}, (13.8.9)
\begin{displaymath}
\s ab=\chi_b\left(\exp\left(-2\pi i\frac{a+\rho}{k+h^\vee}\right)\right).
\end{displaymath}
Here, $\chi_a$ is the character of the representation of
$G$ with highest weight $a$.

Let us check that the two formulas agree (this is essentially
the solution to Exercise 13.35 of \cite{Ka}).
Let  $w\in W$ and $q\in(k+h^\vee) Q^\vee$ and suppose that both $a$ and
$w*a+q$ are dominant integral weights. Then it is easy to see from
the Weyl character formula (see \cite{Hu})
that if $\lambda\in (k+h^\vee)^{-1}P$,
\begin{displaymath}
\chi_{w*a+q}(\exp(2\pi i\lambda))=\epsilon(w)\chi_a(
\exp(2\pi i \lambda)).
\end{displaymath}
There is
a unique element in each affine Weyl group orbit in the
shifted Weyl alc\^ove $I_k+\rho$.
Using these facts and the formula for the multiplicities
in the decomposition of tensor products $\chi_b\chi_c=\Sigma
n^a_{bc}\chi_a$, we deduce \Ref{sss}.

Let us apply this to our example. Identify $\h^*$ with
$\C^N/\C(1,1,\dots,1)$. Then integral weights are
classes $a=[a_1,\dots,a_n]$ of $n$-tuples of integers defined
modulo $\Z(1,\dots,1)$.
The Weyl group $S_N$ acts in the obvious way, and
a weight is dominant if $a_j\geq a_{j+1}$. The affine
reflection $s_0$ is
\begin{equation}\label{s0}
s_0[a_1,\dots,a_N]=[a_N+k+N,a_2,\dots,a_{N-1},a_1-k-N],
\end{equation}
and $\rho=[N-1,N-2,\dots,0]$. Let
$c=[r,0,\dots,0]$ be the highest weight of $S^r\C^N$. Then
the decomposition rules of tensor products say that
$\No^{a}_{bc}=1$ if $a_j=b_j+l_j$ ($1\leq j\leq N$)
for some integers $l_j$ such that
$0\leq l_j\leq a_{j-1}-a_{j}$, $2\leq j\leq N$
and $\Sigma_jl_j=r$. Otherwise, $\No^a_{bc}=0$.
As $\theta=[1,0,0,\dots,0,-1]$, a dominant weight
$a$ belongs to $I_k$ if and only if $a_1-a_N\leq k$.

We need two properties (see \cite{GW})
of the coefficients $N^a_{bc}$, valid for any
$a$, $b$, $c\in I_k$:
(i) $0\leq N^a_{bc}\leq \No^a_{bc}$, for all $a,b,c\in I_k$,
and (ii) $N^{\sigma(a)}_{\sigma(b),c}=N^a_{bc}$ where
$\sigma([a_1,\dots,a_N])=[k+a_N,a_1,\dots, a_{N-1}]$.
We will also use: (iii) Each orbit of $W_k^\wedge$, acting via $*$
on $\h^*$, contains at
most one point in $I_k$.

Let us now fix $c=[Nl,0,\dots,0]$,  and do the classical
calculation first.

\begin{lemma}\label{le1} Let $c=[Nl,0,\dots,0]$. Then
$\No^a_{ac}=1$ iff $a_j-a_{j+1}\geq l$ for all $j\in\{1,\dots,N-1\}$.
\end{lemma}

\begin{proof} The coefficient $\No^a_{ac}$ is non-zero if and only if
there exist non-negative integers $l_1,\dots,l_N$, summing up to $Nl$,
such that $l_j\leq a_{j-1}-a_j$ if $j\geq 2$
and $[a_1+l_1,\dots,a_N+l_N]=a$. It follows that $l_j=l$ for all $j$,
and this solution obeys the inequality iff $a_{j-1}-a_j\geq l$
for all $j\geq 2$.
\end{proof}\par\medskip

\begin{lemma}\label{le2} Let $c=[Nl,0,\dots,0]$, with $Nl\leq k$
and suppose $a\in I_k$. Then
$N^a_{ac}=0$ if $a_1-a_N>k-l$.
\end{lemma}

\begin{proof}
In this case $\sigma(a)=[k+a_N,a_1,\dots]$, and since
$(k+a_N)-a_1<l$, $\No^{\sigma(a)}_{\sigma(a),c}=0$,
by Lemma \ref{le1}. Therefore $N^a_{ac}=0$, by
properties (i), (ii).
\end{proof}\par\medskip

\begin{lemma}\label{le3} Let $c=[Nl,0,\dots,0]$, with $Nl\leq k$.
Then
$N^a_{ac}=1$ if and only if $a_j-a_{j+1}\geq l$, $1\leq j\leq N-1$,
and $a_1-a_N\leq k-l$
\end{lemma}

\begin{proof} We need to prove only the ``if'' part.
We do this by showing that only the first term in the sum \Ref{sss} is
non-zero. Let us suppose that $a$ obeys the hypothesis of the Lemma,
and that $\No^a_{ac}=\No^{w*a}_{ac}=1$, with $w\neq 1$ and derive a
contradiction.  Since $b=w*a$ is dominant, and is not in $I_k$ by
(iii), we have $b_1-b_N\geq k+1$. Let us choose the representative in
$a$ with $a_N=0$, and identify $a_1,\dots,a_{N-1}$ with the row
lengths of a Young diagram.  Then $b$ is obtained by adding $Nl$ boxes
to this Young diagram, in such a way that $a_i\leq b_i\leq a_{i-1}$.
Then $w^{-1}$ with $w^r{-1}*b=a$ is the unique element mapping $b$ to
$I_k$. This element is constructed as follows: (i) Add, for all $j$,
$N-j$ boxes to the $j$th row of $b$ (this adds $\rho$). (ii) Draw a
vertical line at distance $k+N$ from the end of the $N$th row to the
right of it; the only boxes to the right of this line are in the first
row, and their number is at most $Nl\leq k$.  (iii) Take these boxes
and add them to the $N$th row (i.e., act by $s_0$, see \Ref{s0});
permute the rows to get a Young diagram (i.e.  act by an element of
$W$). (iv) Subtract $N-j$ boxes from the $j$th row, $j=1,\dots,n$.

We obtain in this way a diagram which has $Nl$ boxes more
than the original diagram with row lengths $a_i$ and
whose first row has $b_N+k+1\geq k+1$ boxes. The
two diagrams are equivalent, meaning that the latter
is obtained from the former
by adding the same number of boxes to each row. This
number is at least $k+1-a_1$ which by hypothesis is
strictly larger than $l$. We need thus more than $Nl$ boxes,
and this is a contradiction.
\end{proof}\par\medskip

The dimension of the space of conformal blocks can be now
computed: note that $a\mapsto a-l\rho$ (i.e.
$a_j\mapsto a_j-l(N-j)$) maps bijectively the set of
weights obeying the conditions of Lemma \ref {le3}
onto $I_{k-lN}$ whose cardinality coincides with
the dimension of the space of invariant theta functions
with vanishing condition.

We conclude that the space of invariant theta functions
satisfying our vanishing condition coincides
with the space of conformal blocks, in accordance with
our conjecture.

\section{The Knizhnik--Zamolodchikov--Bernard equations,
and generalized classical Yang--Baxter equation}\label{kZb}

\parag{The KZB equations}\label{kzbe}
The Knizhnik--Zamolodchikov--Bernard (KZB) equations, first
written in \cite{Be2},
 for a holomorphic conformal block
$\omega\in E(U)$ are the horizontality conditions
$\nabla\omega=0$, where $\omega$ is
identified with its image by $\iota_\h$.
To write these equations explicitly, let
us compute the expression of the connection $\nabla$
on $E(U)$ viewed as a subsheaf of $\Vn(U\times\h)$ via
$\iota_\h$.

It is convenient to introduce functions $\rho$, $\sigma_w$, $w\in\C$
expressed in terms of the
function $\theta_1$:
\begin{eqnarray*}
\rho(t,\tau)&=&\partial_t\log\theta_1(t|\tau),\\
\sigma_w(t,\tau)&=&
\frac{\theta_1(w-t|\tau)\partial_t\theta_1(0|\tau)}
{\theta_1(w|\tau)\theta_1(t|\tau)}.
\end{eqnarray*}
See Appendix \ref{A} for details on these functions.
We use the notation
\be
\kappa=k+h^\vee,
\ee
and the abbreviation $X_m$ for $X\otimes t^m$. We also identify
$\g$ as a Lie subalgebra of $\Lg$: $X_0=X\in\g$.
Let $C_\alpha=e_\alpha\otimes e_{-\alpha}$ (see \Ref{Cas}).
Then we can write $L_{-1}$ as
\be
L_{-1}=\frac1\kappa \sum_{n=0}^\infty
\left(\sum_{\alpha\in\Delta}e_{\alpha,-n-1}e_{-\alpha,n}+
\sum_\nu h_{\nu,-n-1}h_{\nu,n}\right).
\ee
Now let $U\subset\co$, and $\omega\in E(U)$, which
we identify via $\iota_\h$ with a function on
$U$ with values in $\Vn$. We then have, for
 fixed $u\in\Vn$,
\be
\kappa\langle\nabla_{z_j}\omega,u\rangle=
\kappa\frac{\partial}{\partial z_j}
\langle\omega,u\rangle-\langle\omega,
(\sum_\nu h_{\nu,-1}h_{\nu}+
\sum_\alpha e_{\alpha,-1}e_{-\alpha})u\rangle.
\ee
Recall that vectors in $V$ are annihilated by
$X_n$, with $X\in\g$, $n>0$.
We now use the invariance of $\omega$ under the action of
$\Lie$.
The functions $t\mapsto
e_\alpha\,\sigma_{\alpha(\lambda)}(t-z_j)$
 are elements of
$\Lie(z,\tau,\lambda)$. They have simple poles at $t=z_j$ with
residue $e_\alpha$.
 As a consequence of the invariance of $\omega$, we have
\be
\langle\omega,e_{\alpha,-1}^{(j)} u\rangle=
\rho(\alpha(\lambda))
\langle\omega,e_\alpha^{(j)} u\rangle -
\sum_{k:k\neq j}
\sigma_{\alpha(\lambda)}(z_k-z_j)
\langle\omega,e_\alpha^{(k)} u\rangle,
\ee
for all $u\in\bigotimes_j V_j$.
We can use this identity to compute the value of $\omega$ on
vectors $e_{-1}^{(j)}u$.
The
flatness condition $\nabla_{\lambda_\nu}\omega=0$ translates to
\be
 \langle\omega,h_{\nu,-1}^{(j)}u\rangle=
\frac{\partial}{\partial\lambda_\nu}\langle\omega,u\rangle-
\sum_{k:k\neq j}
\rho(z_k-z_j)
\langle\omega, h_\nu^{(k)} u\rangle.
\ee
To compute further we need the commutation relation
$[e_\alpha,e_{-\alpha}]=\sum_\nu\alpha(h_\nu)h_\nu$, that follows
from $([e_\alpha,e_{-\alpha}],h_\nu)
=(e_\alpha,[e_{-\alpha},h_\nu])$.
We therefore obtain the formula
\begin{eqnarray}
\kappa\langle\nabla_{z_j}\omega,u\rangle&=&
\kappa\frac{\partial}{\partial z_j}\langle \Pi\omega,u\rangle-
\sum_\nu
\frac{\partial}{\partial\lambda_\nu}
\langle \Pi\omega,h_\nu^{(j)}u\rangle \nonumber \\
& &-\sum_{k\neq j} \langle \Pi\omega,
\Omega^{(k,j)}(z_k-z_j,\tau,\lambda)u\rangle
\label{eq:Bernard}
\end{eqnarray}
where $\Pi=\Pi(\lambda,\tau)$ is (essentially)
the ``Weyl--Kac denominator''
(for any choice of positive roots $\Delta_+$)
\be
q^{\frac{\operatorname{dim}\g}{24}}\prod_{\alpha\in\Delta_+}
(e^{i\pi\alpha(\lambda)}-e^{-\pi i\alpha(\lambda)})
\prod_{n=1}^{\infty}
[
(1-q^n)^{\rank\g}\prod_{\alpha\in\Delta}
(1-q^ne^{2\pi i\alpha(\lambda))})
],
\ee
$(q=e^{2\pi i\tau})$,
and with the abbreviation
\begin{equation}\label{ome}
\Omega(t,\tau,\lambda)=\rho(t)C_0+\sum_{\alpha\in\Delta}
\sigma_{\alpha(\lambda)}(t)C_\alpha.
\end{equation}
We also use the standard notation $\Omega^{(i,j)}$ to
denote $\sum_s X_s^{(i)}Y_s^{(j)}$, if $\Omega=\sum_sX_s\otimes Y_s$.
This notation will be used below also in the case $i=j$.
The $\lambda$ independent factors in $\Pi$ do not play a
role here, but will provide some simplifications later. In deriving
\Ref{eq:Bernard}, we have used that, by the classical product
formula for Jacobi theta functions, $\Pi$ is, up
to a $\lambda$ independent factor, the product
$\Pi_{\alpha\in\Delta_+}\theta_1(\alpha(\lambda))$.
Before continuing, we can use the formula \Ref{eq:Bernard} to
complete the proof of Prop.\ \ref{flat}.

\vs
\noindent {\em End of the proof of Prop.\ \ref{flat}.}
What is left to prove is that $[\nabla_\tau,\nabla_{z_1}]$
on $E(U)$. But from the above formula for $\nabla_{z_j}$
it follows that $\Sigma_j\nabla_{z_j}$ vanishes. Indeed
we have $\omega\Sigma_jh_\nu^{(j)}=0$ by $\h$-invariance,
and the other terms cancel by antisymmetry. As $\nabla_\tau$
preserves conformal blocks, we have $[\nabla_\tau,\Sigma_j
\nabla_{z_j}]=0$, and the claim follows from the fact
that $\nabla_\tau$ commutes with $\nabla_{z_j}$ with
$j\neq 1$. $\;\Box$

\vs\noindent
A more involved but similar calculation
 gives a formula for $\nabla_\tau$, also essentially due to Bernard,
which will be given here
without full derivation,

One of the ingredients is Macdonald's (or denominator) identity
(see \cite{Ka})
\be
\Pi(\lambda,\tau)=
\sum_{q\in Q^\vee}
e^{i\pi\tau\frac1{2h^\vee}(\rho+h^\vee q,\rho+h^\vee q)}
\sum_{w\in W}\epsilon(w)e^{2\pi i(\rho+h^\vee q,w\lambda)},
\ee
implying (one form of) Fegan's heat kernel identity
\be
4\pi i h^\vee\partial_\tau\Pi(\lambda,\tau)
=\sum_\nu\partial_{\lambda_\nu}^2\Pi(\lambda,\tau).
\ee
Here, $\rho$ is half the sum of all positive roots of $\g$,
$W$ is the Weyl group, and $\epsilon(w)$ is the sign of $w\in W$.
The (complex) dimension of $\g$ enters the game through the
Freudenthal-de Vries
strange formula $(\rho,\rho)/2h^\vee=\operatorname{dim}\g/24$.

Let us summarize the results. We switch to the more familiar
left action notation, by setting $\langle X\omega,v\rangle=
-\langle\omega, Xv\rangle$ if $X$ is in a Lie algebra and
$\omega$ is in the dual space to $\g$-module.
We also need the following special functions of $t\in\C$, expressed
in terms of $\sigma_w(t)$, $\rho(t)$ and Weierstrass'
elliptic function $\wp$  with periods $1,\tau$.
\begin{eqnarray*}
I(t)&=&\frac12(\rho(t)^2-\wp(t)),\\
J_w(t)&=&\partial_t\sigma_w(t)+(\rho(t)+\rho(w))\sigma_w(t).
\end{eqnarray*}
These functions are regular at $t=0$. Introduce the tensor
\begin{equation}\label{eta}
\Eta(t,\tau,\lambda)=I(t)C_0+\sum_{\alpha\in\Delta}
J_{\alpha(\lambda)}(t)C_\alpha.
\end{equation}

\begin{thm}\label{KZB}
The image $\omega$ by $\iota_\h$ of a horizontal section
of $E(U)$ obeys the KZB equations
\begin{eqnarray*}
\kappa\partial_{z_j}\tilde\omega&=&
-\sum_\nu
h_\nu^{(j)}\partial_{\lambda_\nu}
\tilde\omega
+\sum_{l:l\neq j}
\Omega^{(j,l)}(z_j-z_l,\tau,\lambda)\tilde\omega,\\
4\pi i\kappa\partial_\tau\tilde\omega&=&
\sum_\nu\partial_{\lambda_\nu}^2\tilde\omega
+\sum_{j,l}
\Eta^{(j,l)}(z_j-z_l,\tau,\lambda)\tilde\omega,
\end{eqnarray*}
where $\tilde\omega(z,\tau,\lambda)
=\Pi(\tau,\lambda)\omega(z,\tau,\lambda)$,
and $\Omega$, $H$ are the tensors \Ref{ome}, \Ref{eta},
respectively.
\end{thm}

\noindent{\em Remark.} For $n=1$, these equations reduce to
$\partial_{z_1}\tilde\omega=1$, thus $\tilde\omega$ is a $V^*$-valued
function
of $\tau$ and $\lambda$ only, and
\be
4\pi i\kappa{\partial\over\partial\tau}
\tilde\omega=\sum_\nu\frac{\partial^2}{\partial\lambda_\nu^2}
\tilde\omega -\eta_1(\tau)\operatorname{Cas(V)}\tilde\omega
-\sum_{\alpha\in\Delta}\wp(\alpha(\lambda))e_\alpha e_{-\alpha}\tilde
\omega,
\ee
where $\rho(z)=z^{-1}-\eta_1z+O(z^2)$, and Cas($V$) is the
value of the quadratic Casimir element $C^{(1,1)}$ in the
representation $V$. This equation was considered recently
by Etingof and Kirillov \cite{EtKi}, who noticed that if
$\g=sl_N$ and $V^*$ is the symmetric tensor product $S^{lN}\C^N$,
$e_\alpha e_{-\alpha}=l(l+1)\Id$ on the one dimensional
weight zero space of $V^*$, and the equation reduces to the
heat equation associated to the elliptic Calogero--Moser--Sutherland--%
Olshanetsky--Perelomov integrable $N$-body system:
\be
4\pi i\kappa{\partial\over\partial\tau}
\tilde\omega=\sum_\nu\frac{\partial^2}{\partial\lambda_\nu^2}
\tilde\omega -\eta_1(\tau)l(l+1)N(N-1)\tilde\omega
-l(l+1)\sum_{i\neq j}\wp(\lambda_i-\lambda_j)\tilde\omega.
\ee
See also \ref{example} for a description of the space of
conformal blocks in this case.

\parag{The classical Yang--Baxter equation}
The tensor $\Omega^{(1,2)}=\Omega^{(1,2)}(z_1-z_2,\tau,\lambda)
\in \g\otimes\g$ obeys
the ``unitarity'' condition
\be
\Omega^{(1,2)}+ \Omega^{(2,1)}=0.
\ee
Let us remark that the fact that the connection is flat is
then equivalent to the identity
\begin{eqnarray*}
\sum_\nu
\partial_{\lambda_\nu}\Omega^{(1,2)}h_\nu^{(3)}+
\sum_\nu\partial_{\lambda_\nu}\Omega^{(2,3)}h_\nu^{(1)}+
\sum_\nu\partial_{\lambda_\nu}\Omega^{(3,1)}h_\nu^{(2)} & & \\
-[\Omega^{(1,2)},\Omega^{(1,3)}]
-[\Omega^{(1,2)},\Omega^{(2,3)}]
-[\Omega^{(1,3)},\Omega^{(2,3)}] &=& 0
\end{eqnarray*}
in $\g\otimes\g\otimes\g$.
This identity may be thought of as the genus one generalization
of the classical Yang-Baxter equation. It admits an
interesting ``quantization'' \cite{icm}.

\appendix
\section{Lie algebras of meromorphic functions}\label{A}
\noindent We have the following explicit description of
$\gz$.
Let $\rho$, $\sigma_w$, ($w\in\C$)
be meromorphic $\Z$-periodic functions on the complex plane,
whose poles are simple and belong to $L(\tau)$, and such that
\bea
\rho(t+\tau)&=&\rho(t)-2\pi i\\
\sigma_w(t+\tau)&=&e^{2\pi iw}\sigma_w(t)\\
\sigma_w(t)&\sim&{1\over t}\;,\qquad t\to 0.
\eea
Such functions exist (for $w\in\C-L(\tau)$)
and are unique, if we require that $\rho(-t)=-\rho(t)$.
They can be expressed in terms
of the Jacobi theta function $\theta_1$:
\bea
\rho(t)&=&\frac{\partial}{\partial t}{\log}\,\theta_1(t|\tau),\\
\sigma_w(t)&=&
\frac{\theta_1(t-w|\tau)
\theta_1^\prime(0|\tau)}
{\theta_1(t|\tau) \theta_1(-w|\tau)},\\
\theta_1(t|\tau)&=&
-\sum_{n=-\infty}^{\infty}
e^{2\pi i(t+\frac12)(n+\frac12)+\pi i\tau(n+\frac12)^2}.
\eea
Here a prime denotes a derivative with respect to the first argument.

\begin{proposition}\label{expl}
For $\alpha\in\Delta\cup\{0\}$, $\lambda\in\h$,
and $(z,\tau)\in \co$, the meromorphic functions of $t$
(defined as limits at the removable singularities $\alpha(\lambda)\in\Z$)
\bea
X(e^{2\pi i\alpha(\lambda)}-1)\sigma_{\alpha(\lambda)}(t-z_1),\\
X(\sigma_{\alpha(\lambda)}(t-z_l)-\sigma_{\alpha(\lambda)}(t-z_1)),
\qquad 2\leq l\leq n,\\
X\frac{\partial^j}{\partial t^j}\sigma_{\alpha(\lambda)}(t-z_l),\qquad j\geq 1,
\qquad 2\leq l\leq n,
\eea
$X\in\g_\alpha$ or $\h$ if $\alpha=0$,
are well defined provided $|\operatorname{Im}\alpha(\lambda)|<
\operatorname{Im}\tau$ and belong to $\cal L(z_1,\dots,z_n,\lambda,\tau)$.
If $\alpha$ runs over $\Delta\cup\{0\}$ and $X$ runs over a basis of
$\g_\alpha$ ($\h$ if $\alpha=0$), then these functions form a
basis of $\Lie(z,\tau,\lambda)$.
\end{proposition}

\begin{proof} It is easy to check that these
functions belong to $\Lie(z,\tau,\lambda)$.
Let $\Lie^{\leq j}(z,\tau,\lambda)$ be given
by the functions in $\Lie(z,\tau,\lambda)$ whose pole
orders do not exceed $j$. By the Riemann--Roch theorem,
\be
d(j):=\dim(\Lie^{\leq j}(z,\tau,\lambda))=\dim(\g)jn,
\ee
if $j\geq 1$.
Indeed $\Lie^{\leq j}(z,\tau,\lambda)$ is the space of holomorphic
sections of the tensor product of a flat vector bundle
on the elliptic curve by the line bundle associated to $jD$,
where $D$ is the positive divisor $\Sigma z_i$.

The functions given here are linear independent, as can
be easily checked by looking at their poles, and have
the property that for $j\geq 1$, the first $d(j)$ functions
belong to $\Lie^{\leq j}(z,\tau,\lambda)$.
\end{proof}\par\medskip

To obtain a basis outside the strip
$|\operatorname{Im}\alpha(\lambda)|<
\operatorname{Im}\tau$ we can transport our basis using
the following isomorphisms.

\begin{proposition}
Let $(z,\tau)\in\co$, and $q,q'\in P^\vee$.
Then the map sending $X\in\Lie(z,\tau,\lambda)$
to the function
\be
t\mapsto \exp(2\pi it\ad\,q')X(t),
\ee
is a Lie algebra isomorphism from
$\Lie(z,\tau,\lambda)$ to $\Lie(z,\tau,\lambda+q+q'\tau)$.
\end{proposition}

\noindent For any open subset $U$ of $\co$, $\co\times\h$ or
$\co\times G$
define $\Lie^{\leq j}(U)$, $\Lie_\h^{\leq j}(U)$,
$\Lie_G^{\leq j}(U)$ to be the space of functions
in $\Lie(U)$, $\Lie_\h(U)$, $\Lie_G(U)$, respectively,
whose pole orders do not exceed $j$.

\begin{corollary}\label{freeA}
The sheaves $\Lie^{\leq j}$, $\Lie_\h^{\leq j}$ are
locally free, finitely generated for all $j\geq 1$.
Moreover for each $x\in \co\times\h$, every
$X\in\Lie_\h(x)$ extends to a function in $\Lie_\h^{\leq j}(U)$
for some $j$ and $U\ni x$.
\end{corollary}

\noindent The proof in the case of $\Lie^{\leq j}$ is obtained
by setting simply $\lambda=0$.

We wish to extend this result to $\Lie_G$.
Let us first notice that the function $\sigma_w(t)$
is actually a meromorphic function of $e^{2\pi iw}$.
Thus if $g=\exp(2\pi i\lambda)$, the
functions in Prop.\ \ref{expl}  can be
written as $f(\Ad(g),t,z,\tau)X$, where
the meromorphic function $f$ is regular as
a function of the first argument in the range
corresponding to $|\operatorname{Im}\alpha(\lambda)|<
\operatorname{Im}(\tau)$.
Therefore we may extend the definition of the basis
to give a basis of $\Lie_G(z,\tau,g)$ for $g$ in
some  neighborhood of $g=
\exp(2\pi i\lambda)u$, with $\Ad(u)$ unipotent commuting
with $\Ad(g)$ (It is clear that the multipliers are
correct if $g$ is on some Cartan subalgebra, but
such $g$'s form a  dense set in $G$). The pole
structure does not change if the neighborhood
is sufficiently small.
 In this way by choosing properly the
Cartan subalgebra, we find local bases
of $\Lie_G$ in the
neighborhood of all points in $G$ whose semisimple
parts are of the form $\exp(2\pi i\lambda)$ with
$\lambda$ in some Cartan subalgebra and
 $|\operatorname{Im}\alpha(\lambda)|<
\operatorname{Im}(\tau)$, for all $\alpha\in\Delta$.

\begin{proposition}
Let $(z,\tau)\in\co$, and $q,q'\in P^\vee$.
Then the map sending $X\in\Lie_G(z,\tau,g)$
to the function
\be
t\mapsto \exp(2\pi it\ad\,q')X(t),
\ee
is a Lie algebra isomorphism from
$\Lie_G(z,\tau,g)$ to $\Lie_G(z,\tau,\exp(2\pi i(q+\tau q'))g)$.
\end{proposition}

\noindent With the
Jordan decomposition theorem, we get a local
basis around all points of $G$, and we obtain:

\begin{proposition}
The sheaf $\Lie_G^{\leq j}$,is locally free, finitely generated for all
$j\geq 1$.  Moreover for each $x\in \co\times G $, every $X\in\Lie(x)$
extends to a function in $\Lie_G^{\leq j}(U)$ for some $j$ and $U\ni x$.
\end{proposition}

\section{Connections on filtered sheaves}\label{B}

\noindent Let $S$ be a complex manifold, and denote by $\cal O$ the
sheaf of germs of holomorphic sections on $S$. A sheaf of
Lie algebras over $S$ is a sheaf of $\cal O$-modules $\Lie$ with
Lie bracket $\Lie\otimes_{\cal O}\Lie\to\Lie$ a homomorphism
of sheaves of $\cal O$-modules, obeying antisymmetry and
Jacobi axioms. A sheaf of Lie algebras $\Lie$
over $S$ is said to be locally free
if it is  locally free as
$\cal O$-module, i.e., if every $x\in S$ has a neighborhood
$U$ such that, as $O(U)$-module,
 $\Lie(U)\simeq W\otimes\cal O(U)$ for some complex vector space $W$
 In this
case, $\Lie(U)$ is freely generated over $\cal O(U)$
by a basis $e_1,e_2,\dots$ with Lie brackets
$[e_i,e_j]=\Sigma f_{ij}^ke_k$, (with finitely many non-zero
summands) for some holomorphic
functions
$f^k_{ij}$ on $U$.

We will consider the case in which the sheaf $\Lie$ of Lie
algebra is filtered by locally free sheaves of $\cal O$-modules
of finite type. In other words, $\Lie$ admits a filtration
\be
\Lie^{\leq 0}\subset
\cdots\subset\Lie^{\leq j}\subset\Lie^{\leq{j+1}}\subset
\cdots\subset\Lie=\cup_{j=0}^\infty\Lie^{\leq j},
\ee
with $\Lie^{\leq j}$ locally\footnote{i.e., every point of
$S$ has a neighborhood such that the statement holds for the
restriction of the sheaf to this neighborhood}
 isomorphic to some $\C^{n_j}\otimes\cal O$, inclusions
induced from inclusions $\C^{n_j}\subset\C^{n_{j+1}}$,
and such that $[\Lie^{\leq j},\Lie^{\leq l}]\subset\Lie^{\leq j+l}$.
In particular $\Lie$ is locally free.

A sheaf of $\Lie$-modules is a sheaf $V$ of $\cal O$-modules with
an action $\Lie\otimes_{\cal O}V\to V$ which is assumed to be a
homomorphism of $\cal O$-modules. The image sheaf
 of this homomorphism
is denoted by $\Lie V$.
In the filtered situation it is
assumed further that $V$ is filtered by locally free, finitely generated
$\cal O$-modules:
\be
V^{\leq 0}\subset
\cdots\subset V^{\leq j}\subset V^{\leq{j+1}}\subset
\cdots\subset V=\cup_{j=0}^\infty V^{\leq j},
\ee
and that the action is compatible with the filtration,
i.e., $\Lie^{\leq j}V^{\leq l}\subset V^{\leq j+l}$.
In particular $V$ is locally free, and we can define
a dual sheaf $V^*$ locally as
$V^*(U)=\operatorname{Hom}_{\cal O(U)}(V(U),\cal O(U))$.
If $V(U)$ is of the form $\bar V\otimes \cal O(U)$ for some
vector space $\bar V$, then $V^*(U)$ is the space of
functions $u$ on $U$ with values in the dual $\bar V^*$ such
that $\langle u,w\rangle$ is holomorphic for all
$w\in \bar V$. The dual sheaf $V^*$ has a natural structure
of a sheaf of right $\Lie$-modules and we have a natural pairing
$\langle\; ,\;\rangle:V^*\times V\to \cal O$.

We can define the associated
graded objects
\bea
\Gr \Lie&=&\oplus_{j=0}^\infty\Lie^{\leq j}/\Lie^{\leq j-1},\\
\Gr V&=&\oplus_{j=0}^\infty V^{\leq j}/V^{\leq j-1},
\eea
with the understanding that $V^{\leq -1}=0=\Lie^{\leq -1}$.

Then $\Gr \Lie$ is a graded sheaf of Lie algebras acting
on the graded sheaf $\Gr V$ of $\cal O$-modules, and homogeneous
components are locally free and finitely generated.

The sheaf of coinvariants is $V/\Lie V$, and the sheaf
of invariant forms $E$ is locally given by
\be
U\mapsto E(U)=\{\omega\,\in V^*(U)| \omega\,X=0\; \forall X\in\Lie(U)\}.
\ee

In the filtered
situation, $\Lie V$ is filtered, with $(\Lie V)^{\leq j}=
\Sigma_{r+s=j}\Lie_r V_s$ and we have induced homomorphisms
\be
(V/\Lie V)^{\leq 0}\to
\cdots\to (V/\Lie V)^{\leq j}\to (V/\Lie V)^{\leq{j+1}}\to
\cdots\to V/\Lie V=\lim_{\rightarrow} (V/\Lie V)^{\leq j}.
\ee
Locally,
$(V/\Lie V)^{\leq j}(U)$ is  the quotient $V^{\leq j}(U)$
by the submodule of linear combinations of
elements of the form $Xv$, $X\in\Lie^{\leq r}$,
$v\in V^{\leq s}$ with $r+s\leq j$.

A connection $\nabla$ on a sheaf of $\cal O$-modules $V$ is
a $\C$ linear map $V\to \Omega^1\otimes_{\cal O}V$,
where $\Omega^1$ is the sheaf of holomorphic $(1,0)$-%
differential forms on $S$, such that for all open sets
$U\subset S$,
\be
\nabla (fv)=f\nabla v+ df\otimes v,
\ee
for any $f\in\cal O(U)$, $v\in V(U)$.
The notation $\nabla_\xi$ is used to denote the covariant
derivative in the direction of a local holomorphic vector field
$\xi$: if $\nabla v=\Sigma_i \alpha_i\otimes v_i$,
$\nabla_\xi v=\Sigma \alpha_i(\xi)v_i$.
A connection $D$ on a sheaf of Lie algebras is furthermore
assumed to have covariant derivatives being derivations
for all local vector fields $\xi$:
\be
D_\xi[X,Y]=[D_\xi X,Y]+[X,D_\xi Y], \qquad X,Y\in\Lie(U),
\ee
and a connection $\nabla$ on a sheaf of $\Lie$-modules
with connection $D$
is assumed to be compatible with the action, i.e.,
\be
\nabla_\xi(Xv)=(D_\xi X)v+X\nabla_\xi v,\qquad X\in\Lie(U),\quad
v\in V(U).
\ee
Such a connection induces a unique connection, also called
$\nabla$ on $V^*$ such
that for all open $U\subset S$, $u\in V^*(U)$, $v\in V(U)$,
\be
d\langle u,v\rangle=\langle\nabla u,v\rangle
+\langle u,\nabla v\rangle,
\ee

Let $\nabla$ be a connection on
a sheaf $V$ of $\cal O$-modules. If $V$ is filtered
by free, finitely generated $\cal O$-modules $V^{\leq j}$,
we say that $\nabla$ is
{\em of finite depth}
if there exists an integer $d$ such that
$\nabla V^{\leq j}\subset \Omega^1\otimes V^{\leq j+d}$.
The smallest non-negative such integer will be called
depth of the connection.

\begin{thm}
Let $\Lie$ be a sheaf of Lie algebras and $V$ a sheaf
of $\Lie$-modules over a complex manifold $S$.
Suppose that  $\Lie$ and $V$ have
a filtration by locally free finitely generated
$\cal O$-modules,
and compatible connections
$D$ and $\nabla$  of finite depth. If $\Gr V/\Gr\Lie\Gr V$ has
only finitely many non zero homogeneous summands,
then the sheaf of invariant forms $E$ is
 locally free and finitely generated.
\end{thm}
\begin{proof}
Let $z_0\in S$ and $U$ be a neighborhood of $z_0$, such that
the restriction of $V$ to $U$ is free. Thus there exist
vector spaces $\bar V^{\leq j}$, $\bar V$, such that
\be
V^{\leq j}(U)\simeq\bar V^{\leq j}\otimes\cal O(U)
\qquad
V(U)\simeq\bar V\otimes\cal O(U).
\ee
 The assumption that
 $\Gr V/\Gr\Lie\Gr V$ has vanishing components of
degree $\geq N$ means that if $j\geq N$ and $v\in V^{\leq j}(U)$
we have a decomposition (not necessarily unique)
\begin{equation}
v=v'+Xv'',
\label{decom}
\end{equation}
for some $v'\in V^{\leq j-1}$ and $X\in\Lie(U)$.
By iterating this we see that we can take $v'\in V^{\leq N}$.

The first consequence of this is that the restriction map
$E(U)\to V^{\leq l *}(U)$ is injective for all sufficiently
large $l$.

The second consequence is that we can replace the connection
by a connection which preserves $V^{\leq l *}(U)$ for some
large $l$, and coincides with the given one on the image
of invariant forms. The construction goes as follows.

Let us choose a basis $e_1,e_2,\dots$ of $\bar V$ with the property
that, for all $j$, a basis of $\bar V^{\leq j}$ is obtained by taking
the first dim($\bar V^{\leq j}$) elements of this sequence. View
$\bar V$ as the subspace of constant functions in $V(U)$, and
choose a decomposition \Ref{decom} for all $e_i$:
\be
e_i=e_i'+X_ie_i''.
\ee
with $e_i'\in V^{\leq N}(U)$. Define a new connection
$\tilde\nabla$ by
\be
\tilde\nabla e_i=\nabla e_i'.
\ee
This formula uniquely determines a connection $\tilde\nabla$
on the restriction of $E$ to $U$. The
dual connection on $V^*(U)$, also denoted $\tilde\nabla$ is
defined as usual by $\langle\tilde\nabla\alpha,e_i\rangle
=d\langle\alpha,e_i\rangle-\langle\alpha,\tilde\nabla e_i\rangle$.
By construction, this dual connection coincides with
$\nabla$ on invariant forms, and, if $d$ denotes
the depth of the connection $\nabla$, it
 maps $V^{\leq N+d\,*}(U)$
to itself.


If we introduce local coordinates $t_1,\dots,t_n$ around
$z_0$, with $z_0$ at the origin, we see that we have to solve
the following problem: given a subsheaf $E$ of a finitely
generated free sheaf $F$ on an open neighborhood $U$ of the
origin in $\C^n$, with connection
$\tilde\nabla$ on $F$ preserving $E$, show that there
exists an open set $U'\subset U$ containing $z_0$,
such that $E(U')$ is a
free $\cal O(U')$-module.
Write $F$ as $F_0\otimes\cal O$, for a vector space $F_0$.
We may assume that $U$ is a ball centered at the origin.

\begin{lemma}
Let $\xi$ be the vector field $\sum_i t_i\partial_{t_i}$ on $\C^n$,
and $\tilde\nabla$ be a connection on a free, finitely generated sheaf
of $\cal O$-modules
$F=F_0\otimes{\cal O}$ on a ball $U$ centered at the origin of $\C^n$.
For each $\phi\in F_0$ there is a unique  $\hat\phi\in F(U)$
such that $\hat\phi(0)=\phi$, and $\nabla_\xi\phi=0$.
\end{lemma}

\noindent The proof is more or less standard: the $F_0$-valued holomorphic
function $\hat\phi$ on $U$ is a solution of the system of linear
differential equations
\be
\sum_{i=1}^nt_i\frac{\partial}{\partial t_i}\hat\phi(t)
=\sum_{i=1}^nt_iA_i(t)\hat\phi(t),
\ee
for some holomorphic matrix-valued functions $A_i$,
with initial condition $\hat\phi(0)=\phi$. It
is convenient to rewrite this equation in the form
\be
\frac d{dx}\hat\phi(xt)=B(x,t)\hat\phi(xt),
\qquad B(x,t)=\sum t_iA_i(xt).
\ee
In this form we can apply the standard existence and uniqueness
theorem: the unique solution with initial condition
$\phi$ is given by the absolutely convergent Dyson series
\be
\hat\phi(t)=\phi+\sum_{m=1}^\infty
\int_{\Delta_m}B(x_1,t)\cdots B(x_m,t)\phi\,dx_1\cdots dx_m,
\ee
The domain $\Delta_m$ of integration is the simplex
$0<x_1<\cdots<x_m<1$. It is clear from this formula that
$\hat\phi$ is holomorphic on $U$. This concludes the
proof of the lemma.

Let $E_0$ be the subspace of $F_0$ consisting of all values at $0$
of sections of $E(U')$ where $U'$ runs over all open balls
contained in $U$ and centered at the origin.
Let $e_1,\dots,e_r$ be a basis of $F_0$ such that the
first $s$ $e_i$ build a basis of $E_0$.
The homomorphism of $\cal O(U')$-modules
\bea
\tau: E_0\otimes\cal O(U')&\to& F(U'),\\
\phi\otimes h&\mapsto&\hat \phi h,
\eea
is injective since $\hat\phi$ vanishes if and only if
$\phi$ vanishes. We claim that the image of $\tau$ is precisely
$E(U')$, if $U'$ is small enough.
Let $\psi\in E(U')$.
Then, we can write $\psi$ as
\be
\psi(t)=\sum_{j=1}^ra_j(t)\hat e_j(t),
\ee
for some holomorphic functions $a_j(t)$. By assumption,
$a_j(0)=0$ if $j>s$.
Since $\tilde\nabla\hat e_i=0$, we have
\be
\tilde\nabla\psi(t)=\sum_{j=1}^r\sum_{i=1}^n t_i\partial_{t_i}
a_j(t)\hat e_j(t).
\ee
But $\tilde\nabla$ preserves $E$ and, therefore,
$\Sigma_i\partial_{t_i}a_j(t)=0$ if $j>s$. It follows
that $a_j(t)=a_j(0)=0$ if $j>s$.
We have shown that $E(U')$ is contained in the
image of the homomorphism $\tau$.
Now let,
 for $j=1,\dots,s$, $\psi_j(t)$ be sections of $E(U')$ such
that $\psi_j(0)=e_j$. Such sections exist, by
definition of $E_0$, for some neighborhood $U'$.
Then, the construction above gives
\be
\psi_j(t)=\sum_{l=1}^s a_{jl}(t)\hat e_l(t).
\ee
The holomorphic matrix-valued function $(a_{ij}(t))$ is the unit
matrix
at $t=0$ and is thus invertible for $t\in U'$, if the ball
$U'$ is small enough. We conclude
that $\hat e_j\in E(U')$, which completes the proof.
\end{proof}\par\medskip

Let us see how this applies to our situation, following
\cite{TUY}. For us
$S$ is either of $\co$, $\co\times\h$, $\co\times G$, and $\Lie$ is
the corresponding sheaf of Lie algebras, which we denoted $\Lie$,
$\Lie_\h$, $\Lie_G$, respectively.  The module $V$ is the free graded
$\cal O$-module $\Vhn\otimes\cal O$. The key observation is that
$\Gr(\Lie)_j$ consists of the degree $j$ part of
$(\g\otimes\C[t^{-1}])^n\otimes
\cal O$ for all sufficiently large $j$. Moreover $\Gr(V)=
V$ canonically since $V$ is graded, and the action of elements
of sufficiently high degrees in
$\Gr(\Lie)$ on $\Gr(V)$ comes from the action of
$\g\otimes\C[t^{-1}]$ on the factors $V^\wedge_j$.

The fact that $\Gr V/\Gr\Lie\Gr V$ has only finitely many
non trivial homogeneous components follows
then from the fact that $V^\wedge/t^{-N}\g\otimes\C[t^{-1}]$ is
finite dimensional for all positive integers $N$,
which is proved in \cite{TUY} using Gabber's theorem.

\end{document}